\documentclass[preprint2]{aastex}

\usepackage{graphicx}
\usepackage{color}
\usepackage{xspace}
\usepackage{amsmath}

\newcommand{\be}{\begin {equation}}
\newcommand{\ee}{\end {equation}}
\newcommand{\ltsima}{$\; \buildrel < \over \sim \;$}
\newcommand{\simlt}{\lower.5ex\hbox{\ltsima}}
\newcommand{\gtsima}{$\; \buildrel > \over \sim \;$}
\newcommand{\simgt}{\lower.5ex\hbox{\gtsima}}

\newcommand{\HF}{{H_{fix}}}
\newcommand{\EF}{{E_{fix}}}

\newcommand{\fr}{\it Fr}
\newcommand{\gr}{\kern 2pt\hbox{}^\circ{\kern -2pt K}} 
\def\msun{\rm{\,M_{\odot}}}

\shortauthors{Valdarnini R.}

\begin{document}


\title{
A multifiltering study of turbulence in a large sample of simulated
 galaxy clusters}


\author{R. Valdarnini$^{1,2}$ }
\affil{$^1$SISSA, Via Bonomea 265, I-34136, Trieste, Italy}
\affil{$^2$Iniziativa Specifica QGSKY, Via Valerio 2, I-34127 Trieste, 
Italy}

\email{valda@sissa.it}



\begin{abstract}

   We present results from a large set  of 
 N-body/SPH hydrodynamical cluster simulations aimed at studying the 
statistical properties of turbulence in the ICM. The numerical 
hydrodynamical scheme employs a SPH formulation
in which gradient errors  are strongly reduced by using an  integral approach.
We consider both adiabatic and radiative simulations. 
We construct clusters subsamples according to the cluster dynamical 
status or gas physical modeling, from which we extract 
small-scale turbulent velocities obtained by applying to cluster velocities 
different multiscale filtering methods.  
The velocity power spectra of non-radiative relaxed clusters are mostly 
solenoidal and exhibit  a peak at wavenumbers set by the injection scales  
$\simeq r_{200}/10 $,  at higher  wavenumbers the spectra are steeper  
than Kolgomorov.  Cooling runs are distinguished by much shallower spectra, 
a feature which we interpret as the injection of turbulence at 
small scales due to the interaction of compact cool gas cores with the ICM.
Turbulence in  galaxy clusters is then characterized by multiple 
injection scales, with the small scale driving source acting in addition to
the large scale injection mechanisms.  Cooling runs of relaxed clusters 
 exhibit enstrophy profiles with a power-law behavior over more than two 
decades in radius, and a turbulent-to-thermal energy ratio 
$\simlt 1\%$.  
In accord with  {\it Hitomi} observations, in the core
 of a highly relaxed cluster we find  low level of gas motions.
In addition, the estimated cluster radial profile of the sloshing 
oscillation period is in very good agreement with recent Fornax
measurements, with the associated Froude number satisfying 
$Fr \simlt 0.1$ within  $r / r_{200} \simlt 0.1$.  Our findings suggest 
that in 
cluster cores ICM turbulence approaches a stratified anisotropic regime, 
with weak stirring motions dominated by gravity buoyancy forces  and 
strongly suppressed along the radial direction. 
We conclude that turbulent heating cannot be considered the
main heating source in cluster cores.

\end{abstract}




\section{Introduction}

In the standard hierarchical scenario galaxy 
clusters are the most recent and massive virialized objects 
formed in the Universe.
Gas falling into the dark matter potential during the formation
processes will be heated to virial temperatures ($\sim 10^7 -10^8 \gr$)
and at equilibrium will reside in the form of 
 a fully ionized X-ray emitting  intracluster medium (ICM).
During the process of cluster formation, large scale motions driven 
by merging and accretion processes will generate hydrodynamic instabilities 
 which will inject turbulence into the  ICM. Large eddies 
at the injection scale will form smaller eddies which will transfer energy 
down to the dissipative 
scale, thus heating the ICM. This scenario is supported both 
observationally and numerically  \citep[][and references cited therein]{br15}.

Turbulence in the ICM can be detected either directly using 
high resolution X-ray spectroscopy to measure 
emission-line broadening and thus turbulent velocities, or 
indirectly through a number of effects influencing the physics of the ICM.
Indirect evidence for the presence of turbulence in the ICM has 
been obtained by measuring  the fluctuation spectra of X-ray surface 
brightness maps \citep{su04,ga13},  resonant scattering effects 
\citep{ch04,oz17}, Sunyaev-Zeldovich (SZ) fluctuations \citep{ba12},
and through the diffusion of metals in the ICM \citep{re06}.
The first direct detection of turbulence in the ICM has been provided 
recently   by the  \citet[][hereafter H16]{H16}, who 
measured turbulent velocities of the order of 
 $150~ km s^{-1}$ in the core of the Perseus cluster.

Observational support for the presence of turbulence in the ICM 
favors a low-viscosity or inviscid ICM. This is confirmed by the 
presence \citep{ich17,su17a} of Kelvin-Helmholtz instabilities (KHI) at sloshing 
cold fronts \citep{ma07}. These KHI would otherwise be suppressed 
in a viscous ICM \citep{zu11,ro13}. However, these conclusions may be 
too simplistic. Constraints on ICM viscosity may be affected by projection 
effects in the case of Perseus \citep{zu18}, or by the presence of 
magnetic fields which impact on  the small-scale transport properties 
 of the plasma \citep{sch17,bam18,ba18}.

However, additional support for a low-viscosity ICM comes from
measurements of ICM density fluctuation amplitudes 
\citep{ga13,ek17}, which indicate an ICM with   strongly suppressed 
conduction  with respect to the  Spitzer value.

Turbulent motions are expected to affect ICM properties in a variety
of ways. For instance, the accuracy of cosmological constraints 
extracted from galaxy clusters relies on accurate measurements 
of their gravitating mass. X-ray estimates of cluster masses 
are based on the assumption of spherical symmetry and hydrostatic equilibrium
\citep{ras06,nag07a,pi08,lau09,bi16}.
 However, turbulent motions will provide additional non-thermal 
pressure support which will  bias the hydrostatic equilibrium assumption.

Additionally, non-thermal pressure support  also has a significant effect
on the shape and amplitude of the thermal SZ power spectrum \citep{sh10},
The SZ effect is due to inverse Compton scattering from CMB photons.
Because of its linear dependency on gas density, it can be used to derive 
independent cosmological constraints from SZ cluster surveys 
 \citep{sh10,ba12}.

Other physical processes in the ICM for which the role of turbulence 
is important are the amplification of magnetic fields \citep{do02,ber16,vaz18},
cosmic ray re-acceleration \citep{ek17}, and transport of metals \citep{re06}.

Finally, turbulence in the ICM has been also proposed as a possible 
heating source to solve the so-called cooling flow problem.
The center of relaxed clusters is often characterized by the presence of 
cool dense cores with cooling times $\simlt 1Gyr$, shorter than the age 
of the Universe. This implies radiative losses which will lead to 
an inward motion producing a `cooling flow' \citep{fa94} 
and large mass accretion rates. This is 
not observed \citep{pe03,sa08},  and some heating sources must be operating  
in the cluster cores to regulate the cooling flows.

Various heating models have been proposed  in the literature to 
balance radiative cooling in cluster cores and to solve the
cooling flow problem. 
Possible physical mechanisms include thermal conduction \citep{ya16a}, 
 dynamical friction due to galaxy motions  \citep{ez04,ki07}  
or turbulent diffusion  \citep{ru11}, turbulent  heating
\citep{fu04,de05,zh14a}, sound wave dissipation \citep{zw18}, and 
feedback from active galactic nuclei (AGN). 

In the latter scenario, the ICM is heated by 
    interaction with buoyantly rising bubbles 
due to the injection of jets    launched from the central AGN.
This heating model appears to be very promising, since it is 
energetically viable and  is supported by X-ray observations of bubbles or 
cavities \citep{fa12}.
However, the physical processes for which the jet mechanical energy 
 is transferred to the ICM thermal energy are not yet well understood
and considerable effort has gone into investigating 
\citep[][and references cited therein]{so16} how the ICM is thermalized
in the proposed scenario.

Numerical simulations are a necessary tool for investigating self-consistently 
 the hydrodynamical flows that take place during
merging and accretion processes driving the  ICM evolution  and, in turn,
the generation of turbulence.
The role of turbulence  in 
hydrodynamical simulations of galaxy
clusters has been investigated by many authors 
\citep{fu04,do05,ia08,ma09,va09a,va11,iap11,vaz12,mi14,mi15,sch16,iap17,sch17,vaz17,wi17}.

A critical issue when analyzing simulation results is the separation of
small-scale chaotic turbulent motion from large scale coherent 
bulk flows. In this context several strategies have been proposed 
based on the use of  low-pass filters \citep{do05,va11,vaz12,vaz17}, 
subgrid modeling \citep{ma09},  adaptive   Kalman filtering
  \citep{sch16}, and wavelets \citep{sh18}.
In particular, \citet{vaz12} developed an iterative multi-scale filtering 
approach to extract turbulent motions from cluster velocities. We will later 
discuss their method in detail, since it will be applied, with some 
modifications, to the simulations presented here.

All of the simulation papers previously cited have used codes based on 
Eulerian  schemes \citep{st92,sto08,fr00,te02,no05,br14},
 with the exceptions  of \citet{do05} and \citet{va11} 
who employed a Lagrangian smoothed particle hydrodynamics (SPH) code 
\citep{gm77,lu77,hk89}

The SPH code has several advantages, which are  very useful in astrophysics 
problems.
Because of its Lagrangian nature, SPH can naturally follow the development of
large matter concentrations. Moreover, the method is Galilean invariant and 
naturally conserves linear and angular momentum.

However, it is well known that in its standard formulation SPH suffers
 from several difficulties 
\citep[see, for example, ][and references cited therein]{va16}.
A first problem of standard SPH is the difficulty in dealing with steep density
gradients present at the interface of contact discontinuities; this
is the so-called 
 local mixing instability  \citep[LMI:][]{pr08,read10}.

Several variants have been proposed for solving this problem;
 here we follow
the approach of \citet{pr08},  who incorporated
an  artificial conductivity term into the SPH thermal equation.
 This term is aimed at smoothing  thermal 
energy at the borders of contact discontinuities, which is 
 equivalent to adding  a heat diffusion term to the SPH 
equations. 
By introducing this term, it is found \citep{wa08,va12}
that in non-radiative simulations of galaxy clusters,  
  the levels of core entropies are then in  agreement
with those produced using grid codes.

The second problem is related to sampling effects. Because a finite 
number of particles is used to model the fluid, the discretization
implies the presence of zeroth-order errors in the momentum 
equation \citep{read10}. To overcome this problem, among other approaches, 
\citet{ga12} proposed  estimating SPH gradients  by evaluating 
integrals and performing a matrix inversion. This tensor approach 
has been tested in in a variety of hydrodynamical
test cases \citep{ga12,ro15,va16,ca17}, with good results.

In particular, it has been  found  that the  scheme greatly
improves the numerical modeling of subsonic turbulence \citep{va16}. 
This is a crucial issue, since  it implies that the new SPH formulation
can be profitably used in simulations of galaxy clusters aimed at studying 
turbulence.

We have  incorporated this scheme into our SPH code, which has been
used to construct large samples of simulated galaxy clusters. 
The cluster simulations  were constructed according to the zoom-in method in 
which initial conditions for the individual SPH hydrodynamical runs were
extracted from cosmological dark matter simulations of different 
box sizes. Our final samples comprise $\simeq 200$ clusters, with 
a virial mass range spanning about two orders in magnitude, from 
$\simeq 10^{13} \msun$  up to $\simeq 10^{15} \msun$.

For each cluster we ran an  adiabatic gas dynamical simulation as well as
a radiative run in which the physical modeling of the
gas includes radiative cooling, star formation, and energy feedback from 
supernovae.
Finally, we have used the cluster dynamical status to construct 
two cluster subsamples, which are  identified by including  the most
 relaxed and unrelaxed sample clusters.

The simulation suites are then used to study the statistical 
properties of ICM turbulence  by applying a variety of multifiltering 
algorithms to the gas velocities of the simulation samples. 
The comparison between different results is also aimed at identifying 
the optimal filtering strategy, when in the presence of clusters with 
very different dynamical histories.
Additionally, we also studied the turbulent profiles of an 
individual highly relaxed cluster which we identify as a cool-core cluster.

The paper is structured as follows.
In Sect. \ref{sph.sec} we present the numerical method. The construction 
of the set of simulated clusters is described in Sect. \ref{sample.sec}.
In Sect.  \ref{stat.sec} we describe the  methods we use to quantify 
the statistical properties of turbulence, together with the 
multifiltering strategies used to identify turbulent motions.
The results are presented in Sect. \ref{results.sec} and 
our conclusions are summarized in Sect.  \ref{concl.sec}.


\section{Code description }
\label{sph.sec}
This section describes the main features of the adopted hydrodynamical 
numerical scheme,
for a general review of the SPH method see  \citet{pr12}.

\subsection{Basic equations} \label{subsec:method}

In SPH, the hydrodynamic fluid equations are derived from a set of
point particles with mass $m_i$, velocity $\vec v_i$, density $\rho_i$,
and specific entropy $A_i$ \footnote{We use the convention of having 
Latin indices denoting particles and Greek indices denoting 
the spatial dimensions}. We integrate here the entropy per particle, this
is connected to the thermal energy per unit mass $u_i$ via the
 the particle pressure :
$P_i=A_i\rho_i^{\gamma}=(\gamma-1) \rho_i u_i$, where $\gamma=5/3$ for
a mono-atomic gas.  
The SPH density estimator evaluates the density at the particle position 
$\vec r_i$ by summing over neighboring particles $j$
 \begin{equation}
 \rho_i=\sum_j m_j W(|\vec r_{ij}|,h_i),
    \label{rho.eq}
 \end{equation}
where $W(|\vec r_i-\vec r_j|,h_i)$ is a kernel with compact support 
 which is zero for $|\vec r_i-\vec r_j|\geq\zeta h_i$ \citep{pr12}.
Throughout this paper, we will present simulation results obtained using 
 the cubic B-spline $M_4$ kernel, for which $\zeta=2$.

The smoothing length $h_i$  is determined by the implicit equation
 \begin{equation}
h_i=\eta (m_i/\rho_i)^{1/3}~,
  \label{hzeta.eq}
 \end{equation}
so that  $N_{nn}= {4 \pi (2 \eta)^3 }/{3}$ 
is the  number of neighbors   within a radius $2 h_i$.
 Here we solve numerically the equation for the $h_i$  with  
 $N_{nn}=32$.

The Euler equations can then be derived from a  Lagrangian 
\citep{pr12}; and the momentum equation is
   \begin{equation}
  \frac {d \vec v_i}{dt}=-\sum_j m_j \left[
  \frac{P_i}{\Omega_i \rho_i^2}
  \vec \nabla_i W_{ij}(h_i) +\frac{P_j}{\Omega_j \rho_j^2}
   \vec \nabla_i W_{ij}(h_j)
\right]~,
  \label{fsph.eq}
   \end{equation}
where  $\Omega_i$ is defined as 
   \begin{equation}
   \Omega_i=\left[1-\frac{\partial h_i}{\partial \rho_i}
   \sum_k m_k \frac{\partial W_{ik}(h_i)}{\partial h_i}\right]~.
    \label{fh.eq}
   \end{equation}

In the next section we will present the integral method and show how this
equation needs to be modified.

\subsection{The Integral method}
\label{IAD.sec}

A long standing problem of classic SPH has been the presence of 
  zeroth-order errors in gradient estimates due to sampling effects 
\citep{read10}.
These errors impact on the  momentum equation  and degrade code 
performances in subsonic flows \citep{va16}.
This has led many authors to propose  variants of standard SPH, 
see, e.g., \citet{ho15} for an introduction to several of them.

Here we will follow the approach originally proposed by \citet{ga12},
 in which
SPH first-order derivatives are estimated through the use of integrals. 
It has been shown that this approach greatly improves 
 gradient estimates \citep{ga12,ro15,va16,ca17}, thus strongly reducing the
noise   present in the standard formulation. 
We briefly  outline here the essential features of the method.
  
The gradient of a continuous function $f (\vec r )$ can be 
evaluated by first defining the  integral 

   \begin{equation}
 {I(\vec r) }= \int_V \left [ f(\vec r ^{\prime}) -f (\vec r ) \right ]
\vec \Delta  W( |\vec r ^{\prime} - \vec r |, h ) d^3 r^{\prime}~,
    \label{iad.eq}
   \end{equation}

where  $\vec \Delta \equiv(\vec r ^{\prime} - \vec r )$
and $W$ is  a generic spherically symmetric kernel.
A Taylor expansion   of $f(\vec r ^{\prime})$ 
   to first order  can be inverted to give
 

   \begin{equation}
 \vec \nabla_{\alpha}  f =
  \left [ \tau \right ] ^{-1} _{\alpha\beta} I_{\beta}~,  
    \label{iadc.eq}
   \end{equation}

where 

 \begin{equation}
 \tau _{\alpha\beta}=\tau _{\beta \alpha}= \int \Delta_{\alpha} \Delta_{\beta } W  d^3 r^{\prime}~  
\label{iadd.eq}
 \end{equation}

are the elements of the matrix $\mathcal{T}=  \{ \tau\} _{\alpha \beta}$.

We must now translate the continuous version of these equations into
their SPH discrete counterparts.
The integral  (\ref{iad.eq})  becomes 

   \begin{equation}
{I}_{\beta}(i) = \sum_j \frac{m_j}{\rho_j} f_j \Delta_{\beta }^{ji}
  W(r_{ij},h_i)~ ,  
    \label{iadb2.eq}
   \end{equation}

and for the matrix  $\mathcal{T}$ of particle $i$ one has

   \begin{equation}
   \tau _{\alpha\beta}(i)=\sum_j \frac{m_j}{\rho_j} 
\Delta_{\alpha} ^{ji} \Delta_{\beta }^{ji}  W(r_{ij},h_i)~. 
    \label{iade.eq}
   \end{equation}

A key step is the use of expression (\ref{iadb2.eq}) to evaluate the integral 
(\ref{iad.eq}), this is a valid  approximation
 as long as the condition 

\begin{equation}
 \sum_j \frac{m_j}{\rho_j} (\vec r_j-\vec r_i )  W_{ij}  \simeq 0
 \label{sumc.eq}
\end{equation}

is satisfied  with a certain degree of accuracy. 
This is crucial because it is  easily shown \citep{ga12} that the gradient 
approximation (\ref{iadc.eq}) is now antisymmetric in the exchange of the pairs
 $ij$, so that the new scheme  maintains 
exact conservation properties.

The validity of the approximations involving the integral methods 
 has been tested in a variety of 
hydrodynamical problems \citep{ga12,ro15,va16},
showing a  strong decrease of errors in gradient estimates and
leading to a significant improvement in code performance. 

To summarize, the adoption of the integral scheme 
requires the evaluation of the $3\times3$ matrix (\ref{iade.eq} ) 
and its inversion. This is in order to substitute 
in the SPH equations the scalars 
$\left[ \vec \nabla_i W_{ik} \right]_{\alpha} $ with the
 following prescriptions:

 \begin{equation}
 \left [ \nabla_i  W_{ik}(h_i) \right] _{\alpha}  \rightarrow
\sum_{\beta} C_{\alpha\beta } (i) \Delta_{\beta }^{ki}  W(r_{ik},h_i)\equiv 
{\mathcal A}_{\alpha,ik}(h_i)~,  
  \label{iadf.eq}
 \end{equation}
 
and

 \begin{equation}
 \left [ \nabla_i  W_{ik}(h_k) \right] _{\alpha}  \rightarrow
\sum_{\beta} C_{\alpha\beta } (k) \Delta_{\beta }^{ki}  W(r_{ik},h_k)\equiv
{\mathcal {\tilde A}}_{\alpha,ik}(h_k). 
 \label{iadg.eq}
 \end{equation}
 
where $\mathcal{C}=\mathcal{T}^{-1}$.  The momentum equation (\ref{fsph.eq} )
then becomes

  \begin{equation}
 \frac {d \vec v_{i,\alpha}}{dt}=-\sum_j m_j \left[
 \frac{P_i}{\Omega_i \rho_i^2}
{\mathcal A}_{\alpha,ij}(h_i) 
 +\frac{P_j}{\Omega_j \rho_j^2}
{\mathcal {\tilde A}}_{\alpha,ij}(h_j)
\right]~.
  \label{fsphw.eq}
   \end{equation}

%


From now on,  we will refer to the new SPH formulation 
   as  integral SPH (ISPH). Results from simulations obtained 
using the classical gradient formulation will be referred to as 
standard SPH. 
 Throughout this paper the velocity divergence and curl of 
 particles  will  be consistently evaluated 
 using their SPH estimators, but with the gradients now computed 
according to the new scheme:

   \begin{equation}
  (\vec \nabla \cdot \vec v)_i \equiv \theta_i   = \frac{1}{\rho_i} 
 \sum_j  \sum_{\alpha} m_j \left[ (\vec v_j -\vec v_i)_{\alpha}
 {\mathcal A}_{\alpha,ij}(h_i) 
\right]~,
   \label{div.eq}
   \end{equation}

and

   \begin{equation}
  (\vec \nabla \times \vec v)_{i,\alpha}    = \frac{1}{\rho_i} 
 \sum_j  \sum_{\beta,\gamma} m_j \left[\varepsilon_{\alpha\beta\gamma}
 (\vec v_i -\vec v_j)_{\beta} {\mathcal A}_{\gamma,ij}(h_i) 
\right]~,
   \label{curlv.eq}
   \end{equation}
where $\varepsilon_{\alpha\beta\gamma}$ is the Levi-Civita tensor.

Finally, note that it is now common practice 
\citep[][and references cited therein]{be16a} to use  
 Wendland kernels \citep{de12} in SPH simulations with a large number of 
neighbors, say $N_{nn}\simgt300$.

This choice of this kernel function is motivated by the need to avoid 
pairing instability, which is absent in the case of the Wendland functions, 
when using a large neighbor number. The latter regime is necessary in order
to suppress errors in gradient estimates, which, as previously outlined, 
is a shortcoming of standard SPH.

{
However, in a battery of hydrodynamical tests \citep{va16}
it has been found that ISPH by far outperforms standard SPH. 
The zeroth-order errors in the momentum
equations being reduced by many orders of magnitude,
  with the accuracy of the results 
in the regime of subsonic flows which is found  
comparable to that of mesh-based codes.
These results then
demonstrate that with the new method it is not necessary to use  
a large neighbor number, and justify our choice of
using a cubic spline with  $N_{nn}=32$.
}

\subsection{Shocks and artificial viscosity } 
\label{subsec:visco}
An artificial viscosity (AV) term must be incorporated into the  SPH 
 momentum equation  to prevent particle streaming  and convert kinetic energy 
into thermal energy  at shocks.
We adopt here the commonly employed formulation \citep{mo97} based on
 Riemann solvers:

  \begin{equation}
 \frac {d \vec v_{i,\alpha}}{dt}=-\sum_j m_j 
\Pi_{ij}
{\mathcal {\bar A}}_{\alpha,ij} ~,
  \label{dvdt.eq}
   \end{equation}

where 
   \begin{equation}
{\mathcal {\bar A}}_{\alpha,ij}=\frac{1}{2}
 \left[
{\mathcal A}_{\alpha,ij}(h_i) +
{\mathcal {\tilde A}}_{\alpha,ij}(h_j)
\right]
  \label{asym.eq}
   \end{equation}

   and $\Pi_{ij}$ is the AV tensor.
 This  takes  the form
   \begin{equation}
\Pi_{ij} =
 -\frac{\alpha_{ij}}{2} \frac{v^{AV}_{ij} \mu_{ij}} {\rho_{ij}} f_{ij}~,
  \label{pvis.eq}
 \end{equation}
here $\rho_{ij}=(\rho_i+\rho_j)/2$ is the average density, 
  $\mu_{ij}= \vec v_{ij} \cdot
 \vec r_{ij}/|r_{ij}|$ if $ \vec v_{ij} \cdot \vec r_{ij}<0$ but zero
 otherwise,  $\vec v_{ij}= \vec v_i - \vec v_j$ and 
 $\alpha_{ij}=(\alpha_i+\alpha_j)/2$ is the symmetrized AV parameter.
 The signal velocity 
$v^{AV}_{ij}$ is estimated  as
   \begin{equation}
v^{AV}_{ij}= c_i +c_j - 3 \mu_{ij}~,
  \label{vsig.eq}
 \end{equation}
 with $c_i$ being the sound velocity. The  factor 
$f_{ij}=(f_i+f_j)/2$, where 
   \begin{equation}
  f_i=\frac {|\vec \nabla \cdot \vec v|_i}
  {|\vec \nabla \cdot \vec v|_i+|\vec \nabla \times \vec v|_i}~,
   \label{fdamp.eq}
 \end{equation}

is   introduced  in order to limit
 the AV in the presence of shear flows \citep{ba95}.

In modern SPH formulations, in order to reduce the amount of AV away from 
shocks , the parameter $\alpha_i$ is allowed  to change with time.
This approach was first proposed by
 \citet{mm97}; in their scheme  $\alpha_i$ can increase, up to a maximum 
value $\alpha_{max} $, only in the presence of a converging flow 
($  \theta_i <0$)  and quickly decays
to a minimum value $\alpha_{min}$ afterwards.

{ 
Here will follow the   \citet{cul10} scheme, which uses 
 the time derivative of the velocity divergence
($ {\dot \theta}_i $) to discriminate between pre- and post-shock regions.
The former are identified by the condition $\dot \theta_i<0$, 
where $\dot \theta_i$ is evaluated here by interpolating $ \theta_i$ between 
timesteps. We refer to \citet{cul10} for a detailed description of the method.

Finally, in the implementation of this AV scheme within ISPH, 
two considerations are worth noting.  
The first is that, as demonstrated by
\citet{cul10}, it is crucial to use  higher order velocity gradients
 to prevent false shock detection. This requirement is naturally fulfilled
by ISPH, for which the velocity divergence and curl are calculated 
using Equations (\ref{div.eq}) and (\ref {curlv.eq}). 

The other issue concerns the setting of a floor value  for the 
 $\alpha_i$'s.  A minimum value $\alpha_{min}$   for the viscosity parameters
 is required in order to maintain particle order away from shocks 
\citep{mm97}.  From their tests \citet{cul10} argue  that 
post-shock particle reordering is not prevented even  when
 $\alpha_{min}=0$.  This makes the scheme fully inviscid away from shocks, 
but we prefer here to limit velocity noise by setting 
 $\alpha_{min}=0.1$   \citep[see also ][]{wad17}.
Note that in previous hydrodynamical tests \citep{va16} 
we already used  this AV scheme with $\alpha_{min}=0.1$, with good results 
}

%
\subsection{Dissipative terms}
\label{subsec:ac}
The entropy production rate due to dissipative processes,
 both numerical and physical,  
is given by

   \begin{equation}
  \frac {d A_i}{dt} =\frac{\gamma-1}{\rho_i^{\gamma-1}}\{
   Q_{AV} +Q_{AC}+Q_R\}~,
    \label{aen.eq}
   \end{equation}
%
%

where   $Q_{AV}$ is the source term due to numerical viscosity \citep{va16},
and the term $Q_{AC}$ is an artificial conduction (AC)
term  introduced in standard SPH \citep{pr08} to  avoid
inconsistencies  at contact discontinuities.
 This term can be written as
   \begin{equation}
  \left ( \frac {d u_i}{dt} \right)_{AC} =
\sum_j \sum_{\alpha} \frac{m_j v^{AC}_{ij}}{\rho_{ij}}
\left[ \alpha^C_{ij}(u_i-u_j) \right ]  \Delta^{ij} _{\alpha}
   {\mathcal  {\bar A}}_{\alpha,ij}/r_{ij}~,
  \label{duc.eq}
   \end{equation}
where $ v^{AC}_{ij}$ is  the AC signal velocity,
and $\alpha^C_{i}$ is an AC parameter of order unity.
The setting of the AC parameter is detailed in \citet{va12}, for the
 the signal velocity  we adopt the expression  \citep{wa08,va12}
 \begin{equation}
v^{AC}_{ij} = |(\vec v_i-\vec v_j)\cdot \vec r_{ij}|/r_{ij}~.
 \label{vsgv.eq}
 \end{equation}

This choice works well in the presence of gravity \citep{va12}, 
where otherwise   thermal diffusion 
can  otherwise  arise in an equilibrium configuration.

Finally, it is important to stress that this term is important in modeling 
diffusion processes which are absent in standard SPH, the code being purely
Lagrangian. 
In fact , when the AC term is present in the SPH equations,
it is shown  that  
 the galaxy cluster entropy profiles agree well with those found 
using mesh codes \citep{wa08,va12}. 
 
For the cooling runs, the modeling of the gas incorporates 
 radiative cooling, star formation and  energy feedback from supernovae.
  For these simulations the term $Q_{R}= -\Gamma_c(\rho_i,T_i)/\rho_i$ 
accounts for  radiative losses. We refer to \citet{pi08} and \citet{va06} for 
a  detailed description of the  recipes implemented.

\section{Sample construction of simulated clusters}
\label{sample.sec}
The   ensemble of hydrodynamical cluster simulations 
has been constructed by performing a set of 
 individual runs, with initial conditions for each cluster extracted 
from a cosmological N-body simulation with only dark matter.

For the background cosmological model, we assume a flat geometry with the present
matter density $\Omega_\mathrm{m}=0.3$, cosmological constant density
$\Omega_\mathrm{\Lambda}=0.7$, $\Omega_\mathrm{b}=0.0486$, 
and Hubble constant $H_0=70\equiv 100h$\,km\,s$^{-1}$\,Mpc$^{-1}$.
 The scale-invariant power spectrum is normalized to
$\sigma_\mathrm{8}=0.9$ on an
$8 \, h^{-1}$ Mpc scale at the present epoch.

For a given cosmological run, with box size $L_m$,
we identify dark haloes at $z=0$ using a friends-of-friends
 algorithm, so as to detect overdensities in excess of
$\sim200 \Omega_\mathrm{m}^{-0.6}$ within a radius $R_{200}$.
The corresponding mass is defined as $M_{200}$, where
\begin{equation}
  M_{\Delta}= (4 \pi/3) \, \Delta\, \rho_\mathrm{c} \, R_{\Delta}^3
  \label{mdelta.eq}
\end{equation}
denotes the mass contained in a sphere of radius $R_{\Delta}$
with mean density $\Delta$ times the critical density
$\rho_\mathrm{c}(z)=3 H(z)^2/ 8 \pi G$ and
$H(z)=H_0 \left[\Omega_m (1+z)^3+\Omega_{\Lambda} \right]^{1/2}
\equiv H_0 E(z)$.

Dark matter haloes identified in this way  are then sorted
in mass according to their value of $M_{200}$, and the
 $N_m$ most massive are then selected for the hydro runs.
The corresponding  set is denoted as $S_{m}$.

We repeat this procedure four times to generate four samples $S_m$,
which are combined to construct the final cluster sample $S_{all}$.
 We first run an N-body cosmological simulation 
with a comoving box of size $L_{16}=1600h^{-1}$Mpc, to generate 
 a sample $S_{16}$ with $N_{16}$ clusters.
We iterate the whole procedure by halving the box size , 
 $L_{m/2}=L_m/2$,  down to $L_{2}=200h^{-1}$Mpc.
The final sample $S_{all}$ consists of the four samples 
$S_m=\{S_{16},~S_8,~S_4,~S_2\}$, with $N_m=
\{N_{16},~N_8,~N_4,~N_2\}$ clusters.

The number of clusters $N_m$ of sample $S_{m}$  is usually
chosen \citep{bi15}  
 such that the mass $M_{200}$ of the least massive cluster of sample
$S_m$ is greater than the mass $M_{200}$ of the most massive cluster of sample
$S_{m/2}$ . This choice is made  so that the final 
cluster sample $S_{all}$, of a set of cluster masses, reproduces the 
cosmological cluster mass function.

However, our paper here is aimed at the study of ICM turbulence when using
different filtering methods. Therefore, our sample construction is not 
 constrained by cosmological studies and we choose the set of 
values $N_m$ such that we have, for statistical purposes, a fair 
number of massive clusters. 

The values $N_m$ are then
 $N_m= \{N_{16},~N_8,~N_4,~N_2\} = \{28,~10,~33,~120\}$ , 
for a total of $N_{all}=191$ clusters. At $z=0$  the most massive cluster has 
 $M_{200}\simeq 1.7\cdot 10^{15}h^{-1}\msun $ and the
 least massive $M_{200}\simeq 1.8\cdot 10^{13}h^{-1}\msun $; there 
are about $\simeq 35$ clusters with 
  $M_{200}\simgt 5\cdot 10^{14}h^{-1}\msun $.

{ 
The cluster initial conditions for the hydrodinamic simulations 
are found  according to the following zoom-in procedure,
see \citet[][hereafter V11]{va11} for more details.
For each cluster the dark matter particles which at $z=0$ 
are within $r_{200}$  are located back in the original simulation box at the 
initial redshift $z_{in}=49$. A cube of size $L_c \propto M_{200}^{1/3}$ 
enclosing all of these particles is then placed at the cluster center.
A lattice of $N_L=74^3$ grid point is set inside the cube, with a gas
and a dark matter particle associated to each grid node.
Particle positions are then perturbed, using the same random realization
as for the cosmological simulation. Those particles whose positions lie inside
 a sphere of radius $L_c/2$ from the cube center are kept for the 
hydrodynamic simulation. To model the effect of tidal forces, the sphere
is surrounded out to a radius $L_c$ by a shell of dark matter particles.
These particles were extracted from a cube of size $2L_c$ consisting of 
$N_L=74^3$   points and centered as the original cube.

Initially, each cluster is composed  of $\sim 220,000$ gas and
dark matter particles within a sphere of comoving radius $\propto
R_{200}$.  The mass of the gas particles lies in the range between 
$ \sim2 \cdot 10^8\msun $ to $\sim 3\cdot 10^{9} \msun $.
The gravitational softening parameter of the particles $i$ 
scales with the particle mass $m_i$ as  $\varepsilon_i \propto m_i^{1/3}$.
 The relation is normalized by setting
$\varepsilon_i =15~( m_i/6.2\cdot10^8 \msun)^{1/3}$\,kpc. }

A crucial part of our study is a proper identification of the cluster
dynamical state, in order to disentangle the impact on ICM turbulence of 
the level of relaxation. 
We quantify the cluster dynamical state by using,  as a morphological indicator,
the power ratio method \citep{bu95}.
The power ratios are defined as 
$P_r/P_0$, the quantity $P_r$ is proportional to the square of the
$r$-th moments of the projected X-ray surface brightness $\Sigma_X(x,y)$,
 in the plane orthogonal to the line of sight. Here $\Sigma$ is
 measured within a circular aperture of radius $R_{ap}$.

A useful quantity is $\Pi_3(R_\mathrm{ap})=\log_{10} (P_3/P_0)$, 
 which is the first moment  giving an unambiguous detection of
asymmetric structure.
For a fully relaxed configuration, $\Pi_\mathrm{3}\rightarrow - \infty$.
We define $\bar {P}_r$   as the {\it rms} plane average of the moments $P_r$ 
along the three orthogonal lines of sight.
We then evaluate $\bar {\Pi}_3(r_{500})$  at $z=0$ as a 
 cluster dynamical indicator and sort the clusters of sample $S_{all}$ 
according to their values of $\bar {\Pi}_3(r_{500})$.  

We finally identify as dynamically relaxed (RX), or quiescent,
 those clusters for which their
values of  $\bar {\Pi}_3(r_{500})$ are below the threshold value defining
the $25\%$ of the cumulative distribution.
Similarly, those clusters for which $\bar {\Pi}_3(r_{500})$ 
falls among the top $25\%$ of the cumulative distribution are tagged 
as dynamically perturbed (PT). For the subsample RX, the values
of   $\bar {\Pi}_3(r_{500})$ lie in the range $\simeq \left[-9.4, -8\right]$, 
whereas for the perturbed clusters of the  PT subsample 
$\bar {\Pi}_3(r_{500})\simeq \left[-6,-4\right]$.

\section{Statistical measures}
\label{stat.sec}
In this section we describe the implementation of several 
analysis methods which will be used to study the  
statistical properties of the  simulated cluster turbulent 
velocity fields.

\subsection{Power spectrum}
\label{power.sec}

A standard tool used to quantify the properties of homogeneous isotropic
turbulence is the velocity power spectrum $E(k)$. 
This is evaluated  by  computing the discrete Fourier transform 
 $\vec {\tilde u_{w}}^d(\vec k)$  
of the weighted velocity field $ \vec u_{w}(\vec x)\equiv w(\vec x) \vec u(\vec x)$, 
where $w(\vec x)$ is a weighting function which can take the values $w=1$ or
 $w(\vec x)\propto \sqrt {\rho(x)}$, the latter being a natural choice in the 
case of compressible turbulence  \citep{ki09}.

The vector $\vec {\tilde u_{w}}^d(\vec k)$  is obtained as follows.
A cube of size $L_{sp}$ with $N_g^3$ grid points  is placed at the
cluster center, and  in accordance with  the SPH prescription 
the velocity field $\vec u_{w}(\vec x_p)$ is then evaluated at the grid points
$\vec x_p$.  
The discrete transforms $\vec {\tilde u_{w}}^d(\vec k)$  
 are then computed using fast Fourier transforms and  used
to evaluate the spherically averaged discrete power spectrum
$\mathcal{P}^d(k)=< |\vec {\tilde u_{w}}^d(\vec k)|^2>$,
where  $k=|\vec k|$.

\begin{table}[t]
\caption{Summary of the filtering functions used to calculate the
local mean velocities.
 The first column indicates the kind of kernel, see text for more details. 
The second column gives the kernel width, as  calculated by 
Equation (\ref{sigma.eq}), and $D$ is the kernel dimensionality.
}  
\label{kernel.tab}      
\centering                          
\begin{tabular}{ccc}        
\hline\hline                 
{\it kernel }  & $\sigma^2/H^2$ & D \\    
\hline                        
  $M_4$ & $3/40$  & 3 \\      
  {\it TH}  & $1/12$  & 1 \\      
   {\it TSC}  & $1/9$  & 1 \\      
\hline                                   
\end{tabular}
\end{table}

Finally,  a dimensionless velocity power spectrum
is defined as 

\begin{equation}
E_v(k)=\frac{1}{L_{sp} \sigma^2_v}\left [ 2 \pi k^2 \mathcal{P}^d(k)
\left (\frac{L_{sp}}{2\pi} \right )^3 \right ],
\label{pow.eq}
\end{equation}
where $\sigma_v \equiv \sigma_{200}=\sqrt {G M_{200}/r_{200}}$ and the normalization 
has been introduced to consistently compare, as a function of 
${\tilde k}_r\equiv |k|L_{sp}/2\pi$,
   spectra extracted from different clusters and boxes.

Moreover,  we also study separately  
the longitudinal and solenoidal components of the power spectrum \citep{ki09}, 
 $E_v(k)=E_s(k)+E_c(k)$. For doing this , we decompose 
the Fourier transformed velocity  into its shear and compressive parts
in  $\vec k-$space:

\begin{eqnarray} 
\vec {\tilde u}(\vec k)_{shear}&=& \frac{\vec k \times \vec {\tilde u}(\vec k)}  {|\vec k|}~,\\
 \vec {\tilde u}(\vec k)_{comp}&=& \frac{\vec k \cdot \vec {\tilde u}(\vec k)}
  {|\vec k|} \,.
  \label{pvisc.eq}
  \end{eqnarray}

 The choice of the cube side length $L_{sp}$ and the number of grid points 
$N_g^3$ is dictated  by several arguments  which limit the 
possible choices (V11). 
For a  Lagrangian code, such as SPH, about half of the cluster mass 
at $z=0$ is located  within a radius of $\sim r_{200}/3$.
To reduce  resolution effects, the size 
of the cube should then ideally be chosen as small as possible, but
this would miss most of the large-scale modes 
which  drive the cluster merging and accretion processes.
We therefore set  $L_{sp}=r_{200}$ as a compromise between these two 
opposing needs,  the scaling $L_{sp}\propto r_{200}$  allowing
  consistent comparison of velocity spectra extracted from different 
clusters. 

Similarly, the grid spacing $L_{sp}/N_g$  scales inversely with 
the  1D number of grid points $N_g$ and its value is bounded by 
  the smallest values of the gas smoothing lengths $h_i$. 
These are smaller in the cluster core regions, where the 
cluster density is highest, and for the simulation parameters adopted 
here their values in these  regions lie in the range 
$h_i\sim 5-20 $~kpc. 
The constraint on the grid spacing is then satisfied 
 by setting $N_g=128$, with higher values leading to 
 undersampling effects in the estimate of 
SPH variables at the grid points. 
Generally,  the optimal choice is  $N_g \sim 2 N_p^{1/3}$, with 
$N_p$ being the number of SPH particles.

Finally, the procedure described here implicitly assumes 
 periodic boundary conditions for the velocity 
field  within the cube domain.
To compensate for spurious effects due to non-periodicity one
should adopt a zero-padding technique \citep{va09a}. 
However, previous results (V11) showed velocity power spectra, extracted 
 following  the above procedure, in line with those obtained taking 
into account non-periodicity effects \citep{va09a}.
Moreover, tests performed using non-periodic fields showed that errors 
due to the periodicity assumption can be considered negligible \citep{vaz17}.

Additionally, we also investigate the scaling behavior of the second
order velocity  correlation function:

   \begin{equation}
  \mathcal{S}_2(\vec r)\equiv <|u(\vec x+\vec r)-\vec u(\vec x)|^2>.
    \label{spu.eq}
   \end{equation}

In principle, the function $S_2(r)$ should be evaluated by computing 
velocity differences for all  particle pairs of the sample.
In practice, we evaluate  $S_2(r)$ by randomly chopping a subsample
of $N_s (\simeq N_{gas}/10)$  particles.
For each particle $s$ of the subsample, we compute the velocity difference 
  $\Delta \vec u=\vec u(\vec x_i+\vec r_{si})-\vec u(\vec x_s)$
for all  particles $i$ of the sample which satisfy 
$r_{si}\leq r_{200}$.
We then bin  the  quantity $|\Delta \vec u|^2$ in the corresponding radial bin 
and perform final averages at the end.

We consider separately both the transverse and longitudinal component
 structure functions.  These are accordingly 
defined as 
$\Delta u_{\perp}=\Delta \vec u  \times \vec r_{si}/ |\vec r_{si}|$ and
$ \Delta u_{\parallel}=\Delta \vec u  \cdot \vec r_{si}/ |\vec r_{si}|$.
We also define density-weighted  velocity structure functions 
by weighting velocities in the same way  as 
in the case of the power spectrum.

\begin{table}[t]
\caption{ { Main parameters characterizing the sets of filtered velocities.
From left to the right: the filter function, the 
label used in the text to indicate
 the set of roots $\{H_i\}$  which for the procedure under consideration
is found to satisfy  Equation (\ref{vtol.eq}), the main feature of the 
root finding method (see text). In the first column the notations {\it TH} 
and {\it SPH} stand for a top-hat and B-spline filter function, 
respectively.
$^a$ For the {\it TSC}  kernel we use a fixed filtering length 
 $H_{fix} = r_{200}/10$; $^b$ Non-shocks filtering lengths 
$H_{ns}$ are obtained 
by applying to the set of gas velocities the shock masking 
procedure described in Sect. \ref{shockr.sec};
$^c$ The TH filtering procedure is applied to  the set of gas velocities 
extracted from standard SPH runs; $^d$ Filtering lengths $H^{\Delta}_{B}(mw)$ 
are obtained by applying  a B-spline mass-weighted filtering.
}
}  
\label{filter.tab}      
\centering                          
\begin{tabular}{ccc}        
\hline\hline                 
{\it filtering} & {\it kernel }  & {\it root}   \\    
\hline                        
\it {TSC}$^a$ &  $H_{fix}$ & $H=r_{200}/10$    \\  
\it {TH} &  $H^2_{th}$ & $n_f=2$   \\      
\it { TH} &  $H^1_{th}$ & $n_f=1$   \\      
\it { SPH} &  $H^{\Delta}_{B}$ & $\Delta=r_{200}/200$   \\  
\it { SPH} &  $H^{\eta}_{B}$ & $\Delta=0.05H_i$    \\
\it { TH}$^b$ &  $H_{ns}$ & $n_f=1,2$      \\  
\it { TH}$^c$ &  $H_{th}(STD)$ & $n_f=1$      \\  
\it { SPH}$^d$ &  $H^{\Delta}_{B}(mw)$ & $\Delta=r_{200}/200$   \\  
\hline                                   
\end{tabular}
\end{table}

\subsection{Filtering}
\label{filter.sec}

By their very nature, turbulent flows exhibit a complex pattern of 
velocity structures, characterized by the presence of a small-scale
 fluctuating velocity component over a wide range of scales. 
A useful approach for analyzing the 
turbulent velocity field consists of introducing a filtering
procedure  aimed at decomposing the fluid velocity into a large-scale 
 component and a small-scale part \citep{br94,ad00}:

   \begin{equation}
 <\vec u(\vec x,t)>=\int_D G(\vec x -\vec x^{\prime}, H) \vec u
 (\vec x^{\prime}, t) d^3 \vec x^{\prime}~,
    \label{filt.eq}
   \end{equation}

\noindent where $G(\vec x -\vec x^{\prime},H)$ is a  low-pass filtering function
 and $H$ a  filtering scale. 
A local  small-scale turbulent velocity field 
 $\tilde{\vec u}(\vec x)$ is then defined as

   \begin{equation}
   \tilde{\vec u}(\vec x)=\vec u(\vec x)- <\vec u(\vec x)>.
    \label{filtu.eq}
   \end{equation}
 
This decomposition method is commonly referred to as Reynolds
decomposition \citep{ad00}, and constitutes the framework on which
large eddy simulations of turbulence are based  \citep{sch15}.

This filtering approach was first  applied to the study of turbulence in 
galaxy clusters by \citet{do05},
who used a fixed filtering length  in the  range $H\simeq 50-100~kpc$.
However this method can fail in the presence of a cluster with
a complex dynamical status, in which uncorrelated velocity flows 
can coexists with large-scale streaming motions.

These difficulties led \citet{vaz12} to propose the use of a 
 multifilter approach, in which mean velocities are estimated locally
 using an adaptive filtering scheme with a varying filter length $H$.
The  local lengths so found then provide an estimate of the local coherence
scales of the fluid motion.
 
This approach is not unique, for instance \citet{sch16} have recently 
 implemented an adaptive temporal Kalman filter in order 
to extract the random component from the local velocity 
 \citep[see also][]{sch14,sh18}.
The use of different algorithms follows because turbulence is
a non-linear multi scale phenomenon, and in the presence of 
complex flows the definition of a mean velocity \citep{ad00,ka14}
is inherently ambiguous. 
A discussion of this topic is beyond the scope of 
this paper, and we will limit ourselves to the study of turbulence 
in galaxy clusters using the algorithm of \citet{vaz12}.

In their paper, the authors applied the algorithm to the velocity
fields extracted from a set of galaxy clusters simulated using 
the ENZO code. We now describe the  iterative filtering algorithm 
in our case, where galaxy clusters were simulated using a 
Lagrangian SPH code in which fluid elements are represented using
gas particles.

To derive a local mean velocity $\bar {\vec v}$ around each particle
 with position $  \vec x_i$, at each iteration $n$,
 a mean velocity $ \bar {\vec v_i}^n  $, characterized by 
a filtering scale $H_{i}^n$, is computed  as

\begin{equation}
 \bar {\vec v_{i}}^n  =\frac{ \sum_j G(|\vec x_i-\vec x_j|,H_{i}^n) 
  \vec v_j } { \sum_j G(|\vec x_i-\vec x_j|,H_{i}^n) }~,
 \label{vfilter.eq}
 \end{equation}

where, because of the Lagrangian nature of our
hydrodynamical simulations, the  subscript $i$ 
is added  to indicate the eventual dependency of the filtering 
scale  $H_i^n$ and mean velocity $ \bar {\vec v_i}^n$ on the 
particle position $\vec x_i$. The sum is intended over all the 
particles $j$ for which  $|\vec x_i-\vec x_j|\leq H_i^n $.

After having computed $ \bar {\vec v_i}^n $  we define a small-scale 
velocity field as 

\be
\delta v^n_i= v_i-\bar {v_i}^{n}~.
\label{dv.eq}
\ee

If this velocity $\delta v^n_i$ satisfies the convergence criterion
\begin{equation}
 \frac{\delta v^n_i - \delta v^{n-1}_i}{  \delta v^{n-1}_i} \leq \varepsilon~,
 \label{vtol.eq}
 \end{equation}

where $\varepsilon$ is a tolerance parameter, 
the filtering length $H_i^n$ is then the local length scale and 
$\delta v^n_i$ is identified as the local turbulent velocity field.

In all of the considered cases we 
initially set  $\delta v^0_i$ to an arbitrary large value,
 the tolerance parameter to $\varepsilon=0.05$  and 
apply the filtering procedures only 
to those 
gas particles which lie within a cube of side $6 r_{200}$ placed at the
cluster center.
This choice of the tolerance parameter is justified by the
findings of \citet{vaz12}, 
who found convergence in the filtering of velocities for 
$ \varepsilon \simlt 0.1$.

We  considered a suite of filtering procedures, which differs in the choice 
of the filter function $G$ and in the way in which the root length 
 $H_i^n$ is reached.
For the filtering functions, three different shapes have been considered : 
a top-hat function, the B-spline $M_4$ and the 
 triangular-shaped cloud function (TSC) \citep{ho88}.

The top-hat function is the one previously used in studies based on the 
multifilter approach \citep{vaz12,vaz17}, whereas the $M_4$ kernel is 
the B-spline commonly employed in SPH \citep{pr12}.
 Finally, the TSC kernel has been used for the sake 
of comparability with previous works \citep[][V11]{do05}.
 In Equation (\ref{vfilter.eq})
 we set   $G(|\vec x_i-\vec x_j|,H_i^n) =m_j $ 
 for the top-hat and TSC functions, and    
  $G(|\vec x_i-\vec x_j|,H_i^n) =m_j W_{ij}(H^n_i/\zeta) $  for the 
B-spline filter, the latter being the SPH density weighting of particle $j$ at 
point $\vec x_i$.

In order to properly compare the spectral properties of the filtered velocity
 fields, 
we must introduce some comparison criterion between the smoothing properties
of the different filters.
For doing this  \citep{de12},  we compute the kernel standard deviation

   \begin{equation}
   \sigma^2=\frac{1}{D} \int \vec x^2 G(\vec x, H)  d^D \vec x~,
    \label{sigma.eq}
   \end{equation}

where $D$ indicates the dimensionality of the filter under consideration.
The ratio $\sigma/H$ then provides a measure of the filter width, 
which can be used to compare the spectral properties of the
different kernels.

Table \ref{kernel.tab} summarizes the filtering kernels which 
we use, together 
with their $\sigma/H$  ratios. From these we can see that the standard 
deviation of the
different kernels will  be approximately the same if 
 the kernel-support radii satisfy  the equalities

\be
H_B\simeq H_{TH}\simeq H_{TSC}~,
\label{hsigma.eq}
\ee

where $H_B$ refers the B-spline kernel-support radius.  
 It is also useful to relate 
these kernel radii to  the equivalent width of a  Gaussian kernel:

   \begin{equation}
   G(\vec x)=\frac{1}{(2 \pi \sigma^2)^{D/2} } \exp ( - \vec x^2 /2 \sigma^2 )
\equiv {\mathcal N(0,\sigma^2)}~.
    \label{gauss.eq}
   \end{equation}

This connection can be established by means of the B-splines, which 
for large n approach the Gaussian \citep{de12}.
For  the $M_4$ kernel  one has $ M_4 \rightarrow  {\mathcal N(0,H_B^2/12)}$.

%

The filtering schemes which we used, also differ in the way in which the root
$H_i^n$ is reached. For the top-hat function, for which from now on we will 
refer to as TH, 
 we initially subdivide the original cube of size $L_c=6 r_{200}$
 into cells with mesh spacing $H^0=\zeta h_i^{MIN}$, 
where $h_i^{MIN}$ is the minimum value of the gas smoothing lengths $h_i$.

At the $n-th$ iteration the cube is subdivided into $(L_c/H^n)^3$ cells and
particles lying in these cells are then easily identified and tagged 
using a Head-Of-Chain algorithm \citep[HOC,][]{ho88}.
For each particle $i$, a mean velocity $ \bar {\vec v_i}^n$  is then 
computed by summing over all the particles $j$ which satisfy 
  $|\vec x_i-\vec x_j|< H^n $ and  lie inside the cell $\vec q$ of 
particle $i$  or in one of the $26=3^3-1$ neighboring cells.
We denote by $\{ \vec q \}_i^n$ this set of cells.
 If the new mean velocity  $ \bar {\vec v_i}^n$   does not satisfy 
Equation (\ref{vtol.eq}), the whole procedure is then repeated,  increasing 
the filtering length : $H^n \rightarrow H^{n+1} + \Delta H $,
where $\Delta H=f_H r_{200}$ , and $f_H$ is a free parameter.  
Note that we have dropped the dependency of the filtering lengths $H^n$ 
 on the particle $i$, because $H^n$ is the mesh spacing of the cells
 at the $n-th$ iteration, and is common to all the subset of 
particles $i$ which
 have not yet reached convergence.

This is the SPH version of the original multifilter 
algorithm devised by \cite{vaz12}.
However, some minor modifications have been introduced in order to 
exploit the fact that we are using particles instead of cells.
At the step $n$, because of the space partition performed by the
HOC algorithm, there are two different filtering scales which 
can be defined for particle $i$. These are given by the condition

\be
|\vec x_i-\vec x_j|< n_{filter}H^n/2~,
\label{hth.eq}
\ee

where $n_{filter}$ is an integer which can take the values $1$ or $2$. 
We will indicate as $H^1_{th}$  and $H^2_{th}$, respectively, 
the set of root filtering lengths 
$\{H_i\}$ obtained for different values of 
 $n_{filter}$.  When using a $TSC$ filtering function, we do not 
iterate the computation of the velocities $ \bar {\vec v_i}^n$   
and we use a fixed grid with $H_{fix}=r_{200}/10 $.
    
Additionally, we also consider two sets of filtered velocities constructed
using the B-spline kernel. The iterative procedure is similar to the TH cases,
 but here at each iteration we locate neighboring particles using 
 a tree-search method. 
Initially, we set  $H_i^{0}=\zeta h_i $; the sets differ due to the way in 
which the filtering length $H_i^n$ is searched. For those particles which 
at the step $n-1$ fail to satisfy  Equation (\ref{vtol.eq}), 
we increment  $H_i^{n-1}$  according to  two possible 
rules:  either $H_i^{n-1}$   is incremented by a constant 
quantity or by a relative amount:

\begin{equation}
 H_i^{n-1} \mapsto  H_i^n  = \left\{
\begin{array}{lc}
 H_i^{n-1} + \Delta H^n & {\it constant} \\
 H_i^{n-1} (1 + \eta)& \it {relative} ~,  
\end{array}
\right.
 \label{hmap.eq}
 \end{equation}

where for the  constant case we set 
  $\Delta H^n= f_H r_{200}$.   Hereafter, the sets of filtering lengths 
$\{H_i\}$ obtained by these procedures will be indicated   by
  $H^{\Delta}_{B}$   and $H^{\eta}_{B}$, respectively.  
The choice of the root-finding parameters 
$f_H$ and $\eta$ is a critical issue, particularly 
 in those clusters with a complex dynamical status.
It has been found that if the values of these parameters are chosen too large,
 then 
the root finding algorithm could lead to filtering lengths $H_i$ biased toward
high values.
For this reason we set $f_H =5\cdot 10^{-3} $  and 
 in the relative case  $\eta=0.05$.   

Finally, note that all of the root-finding procedures have been performed
 by expressing the filtering lengths in units of $r_{200}$.
This is in order to consistently compare statistical measures 
extracted from different clusters.
The main features of the different sets of filtered velocities are listed in 
Table \ref{filter.tab}.  

\subsection{Shock identifier}
\label{shockr.sec}

The generation of turbulence in galaxy clusters is a process driven by 
accretion flows, falling into the cluster potential well, and inner merger
 events.
Both of these processes produce shocks which in turn generate turbulence.
However, in order to properly study the statistics of turbulent energies it is necessary
 to separate the small-scale random parts of the velocity flows from the shock
components.

This requires the adoption of a shock finding algorithm, and for Eulerian 
methods several schemes have been applied \citep{sk08,scha15,va09b,vaz12}.
The situation is different for SPH simulations, for which  sampling noise 
generically affects shock identification and Mach number estimates.

There have been various methods aimed at detecting shock fronts in SPH 
simulations.
Algorithms based on entropy changes were introduced  by \citet{pf06} and
\citet{ho08}.
{ 
Recently, \citet{be16b} presented a geometrical on-the-fly  shock detector 
which is shown to work well in a variety of test cases.
Here, we will adopt their method  together  with some minor modifications. 
A full derivation of the method is  described in Appendix A.  }

 \begin{figure*}
 \centering
\includegraphics[width=17.2cm,height=8.2cm]{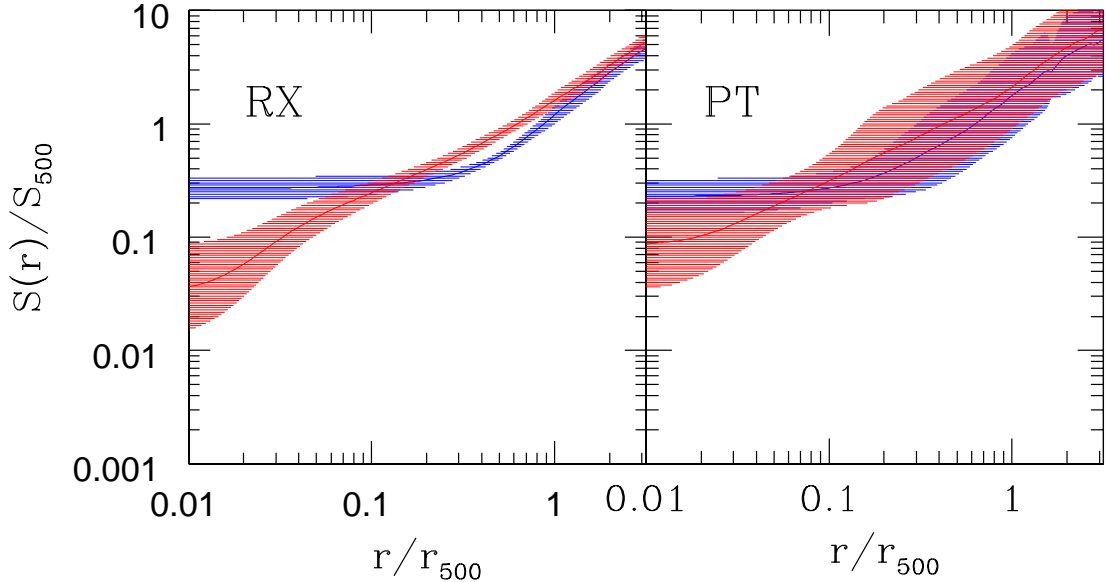}
 \caption{ Ensemble averaged final radial entropy profiles 
are shown for the relaxed (perturbed) subsample in the left (right)
panel. In each panel the entropy profile of the adiabatic (cooling) 
runs is shown in blue (red). The solid lines are the medians and the shaded areas
 represent the limit of the $1\sigma$ dispersion.
 The gas entropy  is defined as $S(r)=k_B T(r)/n_e^{2/3}$ and  is
plotted in units of 
$S_{500}\simeq  1963 
\left[ M_{500}/ ( 10^{15} h^{-1} \msun )\right]^{2/3}keV cm^2$.
}
 \label{figsr.fig}%
 \end{figure*}

Application of the \citet{be16b}  SPH shock finder to the simulations
leads to the identifications of shocks and to the assignement
of individual Mach number $M_i$  to SPH particles.
Shocks identified in this way will be used in some cases to apply a shock
limiting procedure to the TH multifilter algorithm described in 
Sect. \ref{filter.sec}. The  procedure is implemented as follows. 
For a given particle $i$ and filtering scale $H^n$ at the iteration level $n$, 
we stop the iteration if there are within the set of cells
$\{ \vec q \}_i^n$ some particles $s$ for which 
their Mach number $M_s$ is above a certain threshold value $M_{thr}$.
We remove these particles, together with their neighbors, 
 from the cells and define the filtering length $H_i^n$ 
as  the minimum distance between $\vec x_i$ and the remaining particles $j$ in 
the cells: $H_i= \underset{j \in \vec q}{MIN} |\vec x_i-\vec x_j|$.

This shock limiting procedure leads to non-shocks
 filtering lengths $H_i^{\{ns\}}$  which are then smaller than their 
counterparts  $H_i$, obtained without shock masking. 

The choice of the threshold parameter $M_{thr}$ is a
 critical issue since the amount of reduction in the 
 filtering lengths $H_i^{\{ns\}}$, due to the shock limiting 
procedure, depends on the  value of $M_{thr}$.
One has to distinguish between weak shocks ( say $M\sim 1$), 
which can be present in turbulent motions, and strong shocks 
($M\simgt1$) which occur in cluster outskirts, or during merging events, 
 and act as sources of turbulence.
As a compromise, and also for comparative purposes, 
  as in \citet{vaz17} we set here $M_{thr}=1.2$.

\section{Results}
\label{results.sec}
We now apply the statistical methods presented in Sect. \ref{stat.sec} 
to the ICM velocity fields extracted from subsamples of simulated 
galaxy clusters, the analyses being aimed at studying their 
turbulent statistical properties.
For each subsample we extract  statistical  results arising from
considering  velocity fields obtained by applying different 
filtering methods.

\subsection{Global turbulence statistic from cluster subsamples}
\label{global.sec}

As outlined in Sect. \ref{sample.sec}, we perform our statistical analyses 
by constructing two subsamples out of the ensemble of simulated  clusters,
 the subsample membership criterion for the simulated clusters being their
dynamical status. This is identified through the value 
of the power ratio  $\bar {\Pi}_3(r_{500})$ at $r=r_{500}$.
Other choices of dynamical indicators are clearly possible \citep{ra13}, 
but the power ratios  are a commonly employed  
reliable and robust method \citep{we13}.

The relaxed subsample (RX) is defined by those clusters for which their 
  $\bar {\Pi}_3$ values are below the threshold value 
  $\bar {\Pi}^{th}_3\simeq -8 $,  representing the 
 $25\%$ of the cumulative distribution 
  $N_{cl}(<\bar {\Pi}_3)$.   
Similarly, for the perturbed clusters the subsample PT is defined 
by those clusters filling the top  $25\%$
  ($\bar {\Pi}_3 > \bar {\Pi}^{th}_3\simeq -6 $) of the  
  $N_{cl}(<\bar {\Pi}_3)$  distribution.

The threshold values $\bar {\Pi}^{th}_3 $   are chosen with the 
compromise criteria of having  subsample clusters 
with  a well defined cluster dynamical status 
and, at the same time,  subsample  sizes ($N_{cl} \simeq 50)$ 
large enough to allow statistically meaningful comparisons.

In order to assess how realistic the simulations are which we use,  
for the two subsamples we show in Figure \ref{figsr.fig} 
the averaged radial entropy profiles. In each panel we show 
separately the profiles from adiabatic and radiative simulations.
The shaded areas delimit one standard deviation of the subsamples.

We define as entropy the commonly employed related quantity 
 $ S\equiv k_B T/n_e^{2/3}$, where $T$ is the gas temperature 
and $n_e$ the electron density. To allow comparisons with previous 
results, in the plots we show $S(r)$ normalized to
  $S_{500}$. The latter is generically defined according to the 
self-similar model as  \citep{nag07b}


\begin{align}
\lefteqn {S_{\Delta}} & & \simeq  3070 keV cm^2 
\left( \frac{ M_{\Delta} } {10^{15} \msun h^{-1}}\right) ^{2/3} \nonumber\\
& & (\Delta f_b^2)^{-1/3} E(z) ^{-2/3} h^{-4/3}, 
 \label{sdelta.eq}
 \end{align}

where $\Delta=500$, $f_b =\Omega_b/\Omega_m=0.162$  is here the cosmological 
baryon fraction, and for the mean molecular weights  we assume 
$\mu=0.59~,\mu_e=1.14$.

These entropy profiles can be compared with  previous findings 
\citep{ra15,ba17,ha17}.
For instance, a comparison with Figure 1 of \citet{ra15} 
shows  substantial agreement with the radiative entropy profiles  
 shown here.
In their paper, the authors subdivide the sample of simulated clusters 
into cool-core (CC) and non-cool-core (NCC) clusters.
Observationally,  CC clusters are characterized by a dense, cold, compact
core with a cooling time $t_{cr}$ shorter than $H_0^{-1}$. A key feature
of these clusters is that of being associated with a regular X-ray 
morphology. On the contrary,  for NCC clusters   a 
specific feature is an high level of central entropy and a nearly 
flat core entropy profile. These are often associated with a disturbed
morphology.

Various criteria have been proposed to classify CC clusters 
\citep{cav09,mc13}; among these  there is the requirement of having a
central entropy $S_0$ below a threshold value: 
$S_0 < 60 KeV cm^2$. This criterion is also used in \citet{ra15} to
identify simulated CC clusters.
For the cooling runs of the RX subsample, only $4$ clusters out of  
 $\simeq 50 $ have a central entropy $S_0$ above the threshold value.
These results confirm the use of a morphological criterion as a 
CC indicator, as well as the validity of the simulations presented here.

\subsubsection{Filtering lengths} 
\label{subsec:clfilters}

\begin{figure*}[!ht]
 \hspace*{-1.0cm}
\includegraphics[width=18.2cm,height=8.2cm,scale=0.8]{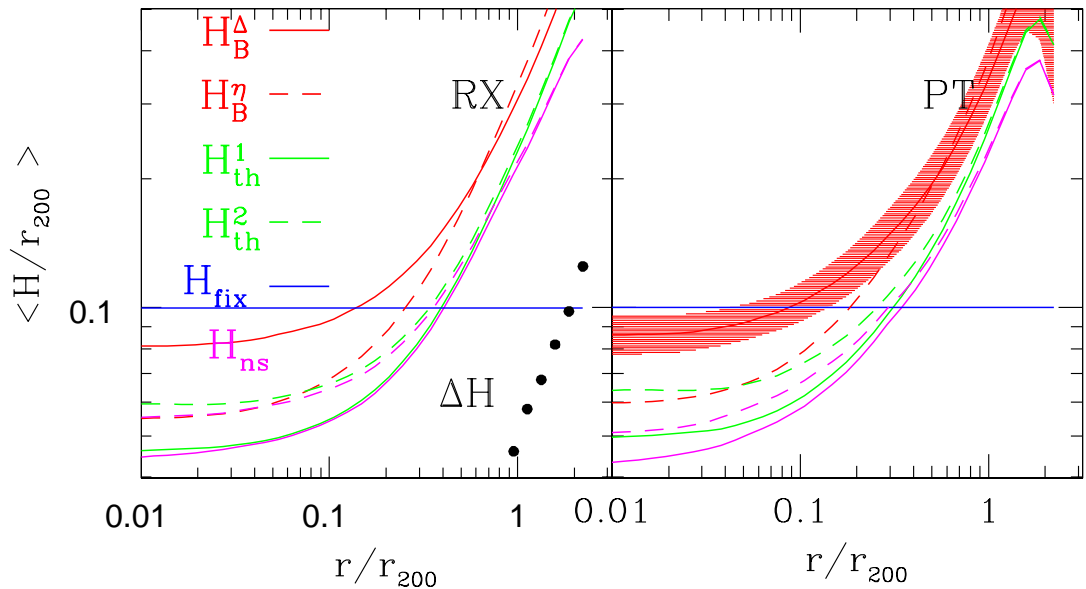}
 \caption{Average radial profiles of the turbulent filtering scales 
$H(r)$ are shown at $z=0$ as a function of $r$ in units of $r_{200}$.
The left panel refers to averaged lengths extracted from the relaxed (RX) 
subsample of adiabatic simulations, the right panel to the perturbed (PT) one.
Different color-codings and line styles indicate curves obtained 
according to the different procedures described in Sect. \ref{filter.sec}.
{ 
The line in blue ($\HF$) refers to the TSC filtering, for which 
we use a fixed filtering length ${\HF}=r_{200}/10$.
The curves in magenta ($H_{ns}$) refer to the corresponding 
  $H_{th}$ curves in green, but are obtained by applying 
the shock-masking procedure of Sect. \ref{shockr.sec}.}
For illustrative purposes, in the right panel the shaded
area indicates  the limit of the $1\sigma$ dispersion of the 
$H^{\Delta}_B$ curve.
{
The black circles in the left panel show the relative difference 
 $\Delta H/r_{200}=(H_{th}-H_{th}$(HR)$)/r_{200}$ between  two 
TH profiles  extracted from the same test cluster.
The profile $H_{th}$ refers to the baseline run and $H_{th}$(HR) 
to a high resolution run (see text).}
}
 \label{fighl.fig}%
 \end{figure*}

\begin{figure*}[!ht]
 \centering
\includegraphics[width=17.2cm,height=13.2cm]{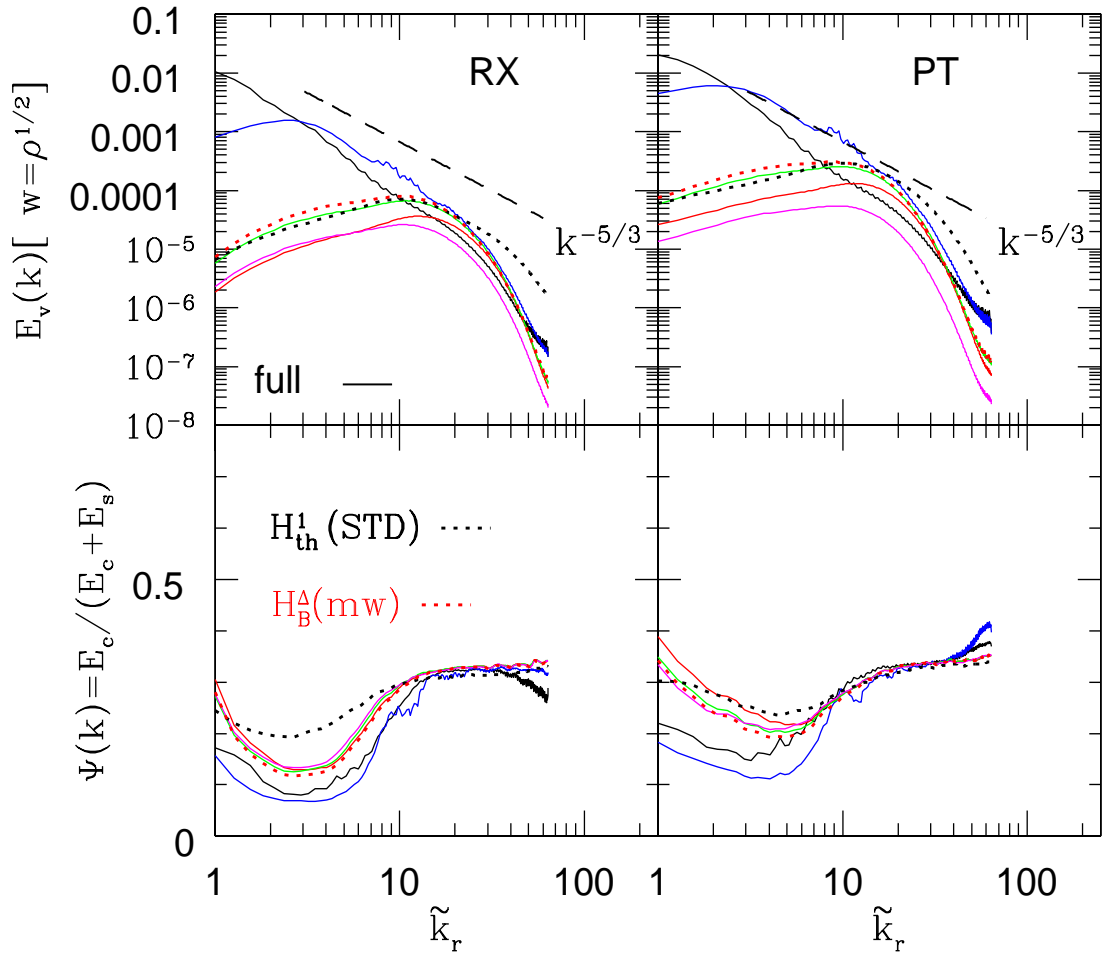}
\caption{The ensemble averaged
  density-weighted velocity power 
spectra $E_v=E_s(k)+E_c(k)$ 
are shown in the top panels as functions of the dimensionless wavenumber 
${\tilde k}_r\equiv kL_{sp}/2\pi$, where $k=|\vec k|$.
The spectra are extracted from the adiabatic cluster simulations 
at $z=0$  using a cube of size $L_{sp}=r_{200}$ with
$N_g^3=128^3$ grid points and are shown up to the wavenumber 
${\tilde k}=N_g/2$. The left (right) panel refers to averages obtained from
 the relaxed (perturbed) subsample RX (PT), see text.
 In each panel the line style and color-coding  is the
same as in Figure \ref{fighl.fig}. The spectra are normalized according to 
Equation  (\ref{pow.eq}).  
The solid black line shows the spectrum of the full (unfiltered)  velocity 
field,  whilst  the dashed  black line indicates the Kolgomorov
scaling. The black dotted line $H_{th}(STD)$ refers to the 
spectra filtered according to  $H^1_{th}$, but extracted from  
standard SPH runs.  The red dotted line is for spectra extracted 
from velocities filtered using the $H_B^{\Delta}$ prescription, but 
mass-weighted (see text).
The bottom panels show the ratio of the longitudinal to total velocity 
power spectra for the same spectra shown in the bottom panels.
}
 \label{figpwavw.fig}%
   \end{figure*}

We have  applied the multifiltering methods 
described in Sect. \ref{filter.sec}
to the ensemble of simulated clusters,  in order to extract different sets
 of filtered velocity fields.
For each filtering procedure, the radial behavior at the final epoch 
of the averaged root filtering lengths $<H(r)>$ is shown separately 
 in Figure \ref{fighl.fig} for each cluster subsample. 
We show there only averages extracted from adiabatic simulations, the 
 $<H(r)>$  profiles of the cooling runs being quite similar.
All of the averaged lengths have been rescaled in units of $r_{200}$; 
 the different curves are labeled according to Table \ref{filter.tab}.
The ensemble average power spectra of the corresponding filtered 
velocity fields are shown in Figs. \ref{figpwavw.fig} 
to \ref{figpwcrw.fig}.

The radial dependence of the different 
 $<H(r)>$  profiles depends on a number of issues related to the 
adopted procedure. In the case of B-spline filtering the behavior 
of  $ < H^{\Delta}_{B}(r) >$ is different from that of 
 $< H^{\eta}_{B}(r) > $.
Specifically, from the left panel of Figure \ref{fighl.fig} 
one sees that the ratio 
 $ < H^{\Delta}_{B}(r) >/  < H^{\eta}_{B}(r) > $ is $\simgt1$ 
for $r/r_{200} \simlt 0.8$
and becomes smaller than unity at larger radii.

This behavior can be understood in terms of the different root finding methods 
 adopted by the two procedures. In both methods the $H^0_i$ are initially 
set to $\zeta h_i$, however their increment $\Delta H$ at each iteration 
is different. The increment ratio between the  $\Delta H$  of the
two methods is given by 
 $ r_H(r)=   \Delta H^{\Delta}_B / \Delta H^{\eta}_B  =f_H r_{200} /
( \eta \zeta h_i(r) ) \propto \rho^{1/3}$, 
 so that it tends to higher values with decreasing radii.  
For the chosen root finding parameters, $r_H(0) \simeq 4$, with
 $r_H(r)\simlt1$ at large radii because of the drop in density.
The corresponding ratio between the root lengths is 
 $  H^{\Delta}_{B} /   H^{\eta}_{B} \simeq 
(1+\eta r_H)/(1+\eta) $, so that  $  H^{\Delta}_{B} $ is about  $30\%$ 
higher  than $H^{\eta}_{B}$   in the cluster cores.

These results show how the set of root values $\{H_i\}$  found using 
the multifilter method depends critically on the set of chosen  
initial values $H_i^{0} $ as well as on the step lengths 
 $  \Delta H $.
In the $H_{B}$  cases, both methods start by setting 
 $H^0=\zeta h_i$, and because in cluster cores the velocity field is
very regular, the root values $\{H_i\}$  are found  at the second iteration.
The difference between the roots is then given by the different
 increments $  \Delta H $ used in the two procedures.

We now examine the radial behavior of the TH filtering lengths. At variance 
with B-spline filtering, here we set the initial grid spacing to a very 
small value $H^0=\zeta h_i^{MIN}$. This guarantees that root finding starts 
 from  mesh values $H$ safely  below those of the generic  root. 
For the same reason we set the grid increment to very small values : 
 $\Delta H=f_H r_{200}$ , with  $f_H = 1/200$.  
From Figure  \ref{fighl.fig}  one can  see  that the average value of 
the root set $H^1_{th}$  is very close to that of 
 $H^{\eta}_{B}$   for $r/r_{200} \simlt 0.1$.
This occurs because the small increments in grid spacing ensure that the 
root values are bracketed without overstepping, whilst for the same reason
 the $ H^{\Delta}_{B} $  are found biased toward high values.

Note that in Figure  \ref{fighl.fig} the $H_{B}$ profiles have been rescaled
 by a factor of two with respect the  corresponding root values; 
this is because according
to the definitions of Sect. \ref{filter.sec} the  $H_{B}$ root is a 
  radius, while for TH filtering the  root length is the grid size.

The averaged profile of the $H^2_{th}$   roots is quite similar to that 
of $H^1_{th}$. This is not surprising since for the filtering procedure 
of  $H^2_{th}$ the search radius is twice that of $H^1_{th}$, 
but the very small value of $ \Delta H $  implies that the two procedures 
 converge to the same root.
In the following parts of this paper,
 this filtering case  will not be discussed 
any more, and  we will refer only to $H^1_{th}$.

For the TH filtering, we have also applied the shock limiting procedure 
described in Sect.  \ref{shockr.sec} to extract from the simulated samples 
a set of roots $H^{(ns)}_{th}$.
As discussed in Sect. \ref{shockr.sec}, the application of a shock limiter 
 leads to filtering lengths $H^{(ns)}_{th}$ smaller than or equal to 
 their counterpart $H_{th}$.
 The average radial profile  $H^{(ns)}_{th}$ depicted 
in Figure \ref{fighl.fig} shows at small radii ($r \simlt r_{200} $) 
a behavior very close to that of the unmasked 
 filtering lengths  $H_{th}$. Beyond $r_{200}$  the $H^{(ns)}_{th}$  
begins to drop low,  because of accretion shocks present in the cluster
outskirts. We thus expect  for relaxed clusters at late epochs the 
effects of masking to be negligible for $r \simlt r_{200} $. 

For the perturbed clusters the average radial profiles 
are displayed in the right panel of Figure \ref{fighl.fig}.
These profiles have been extracted by applying 
the same filter procedures  used for relaxed clusters.
The radial dependence of the profiles is the same as for the RX 
subsample, but the dispersion is higher.
For the sake of clarity in one case ($ H^{\Delta}_{B} $), 
we show the area delimiting the one sigma dispersion.

At $r \simgt r_{200} $,  the filtering lengths show evidence 
of some degree of decrease as the radius increases. This  
indicates that the outskirts of some clusters of the PT subsample
are dynamically unrelaxed, with perturbed velocity flows.

{   
To demonstrate that the $H(r)$ profiles presented here are not affected by
numerical resolution, for a single test cluster  we show in the left
panel of Figure \ref{fighl.fig}  the relative difference 
$\Delta H /r_{200}$ between two distinct TH profiles $H_{th}$.
This is a highly relaxed cluster whose properties are discussed in great
 detail in Sect. \ref{singlecl.sec}. To assess the effects of numerical 
resolution, for this test cluster we extracted the profile  $H_{th}$(HR)
from a high resolution run (HR).  This was performed by running a simulation
 with about twice the 
number of particles of the baseline run.
We then contrast in Figure \ref{fighl.fig}   (black circles) the
difference $\Delta H=H_{th}-H_{th}$(HR) between the baseline and 
HR profile.
The results indicate a relative difference 
$\Delta H /r_{200}\simlt 4 \cdot 10^{-2} $  at radii 
$ r \simlt r_{200}$, thus validating the effectiveness of the adopted 
numerical resolution for the baseline runs.

For the same test cluster, we obtain in Sect.  \ref{singlecl.sec} similar
results when discussing the dependency of the velocity power spectrum
on numerical resolution.
We argue in Sect.  \ref{singlecl.sec} that the weak resolution 
dependency of the ISPH scheme, when compared against 
standard SPH, is a consequence of its ability to suppress gradient errors.
In the standard SPH formulation these errors are strongly affected by
numerical resolution, so that in the new scheme resolution dependency 
 is now subdominant \citep{va16}.

Finally, it is important to emphasize that the applicability of the 
multifilter method requires a well defined separation 
between the coherence scale of bulk flows and that of small-scale motions.
  This in order 
to allow a proper definition of a local mean field.
This condition might not be fulfilled in the case of cluster mergers,
for which the largest injection scales of turbulence could approach
that of large scale motions.

To validate their method, \citet{vaz12} analysed turbulent velocity fields
extracted from a set of idealized test cases. In particular, for
cluster mergers the spectral behavior of the ICM velocity field is found
to be dominated by turbulent motions at spatial scales 
$r\simlt 0.1 -0.3r_{200}$. At larger scales the motion is 
mostly laminar. This is in accord with the results presented in the 
next Section and supports the use of the multifiltering approach
to detect turbulent motions in the ICM.
}

\subsubsection{Power spectra and velocity structures (adiabatic simulations)} 
\label{subsec:clpowa}

\begin{figure*}[!ht]
 \centering
\includegraphics[width=17.2cm,height=13.2cm]{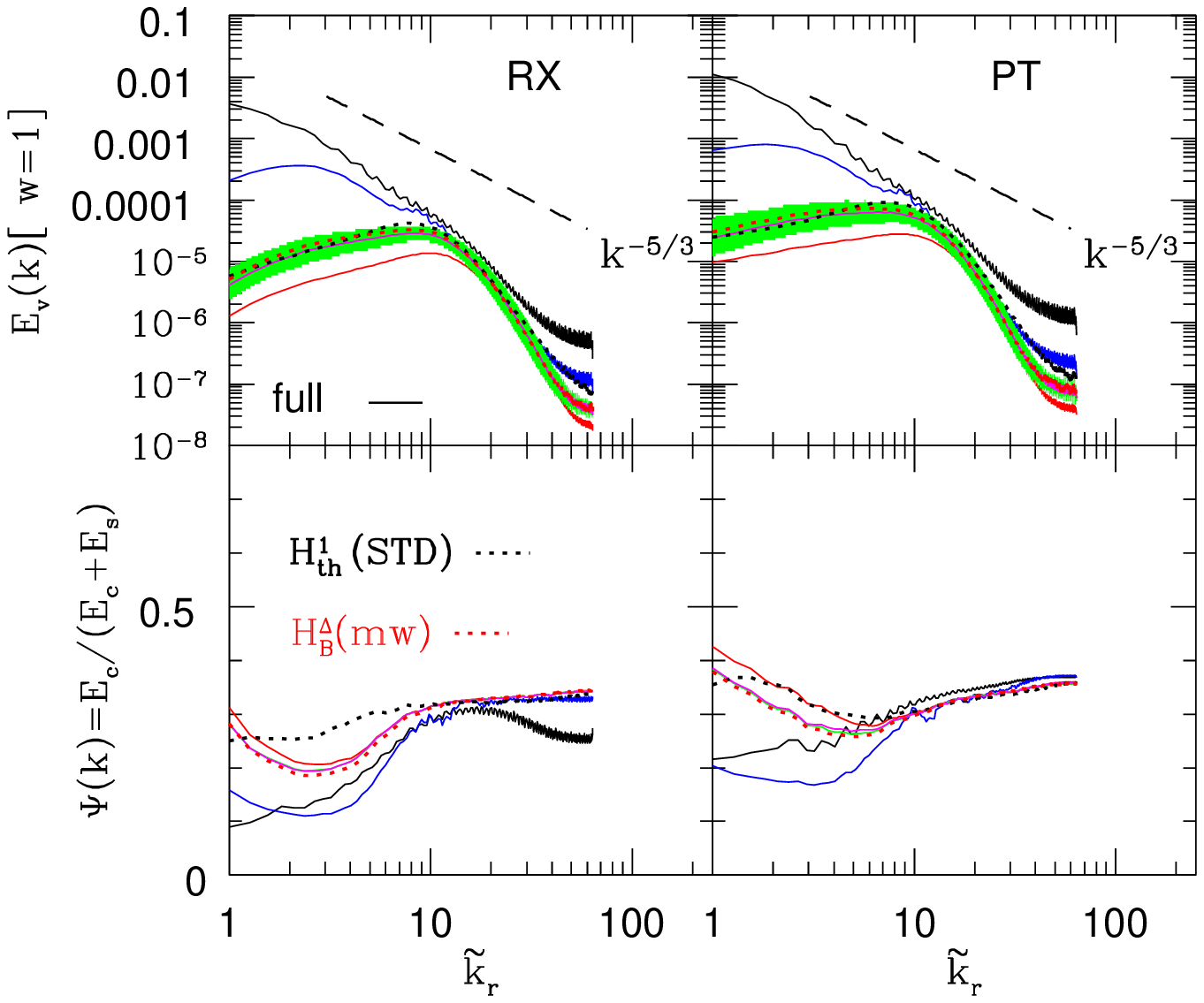}
 \caption{As in Figure \ref{figpwavw.fig}, but spectra are volume weighted.
{ 
The green areas in the top  panels  represent 
 the $1\sigma$ dispersion range of the averaged power spectra,
 filtered according to  the $H^1_{th}$ procedure.
}
}
 \label{figpwav.fig}%
  \end{figure*}

\begin{figure*}[!ht]
 \centering
\includegraphics[width=17.2cm,height=13.2cm]{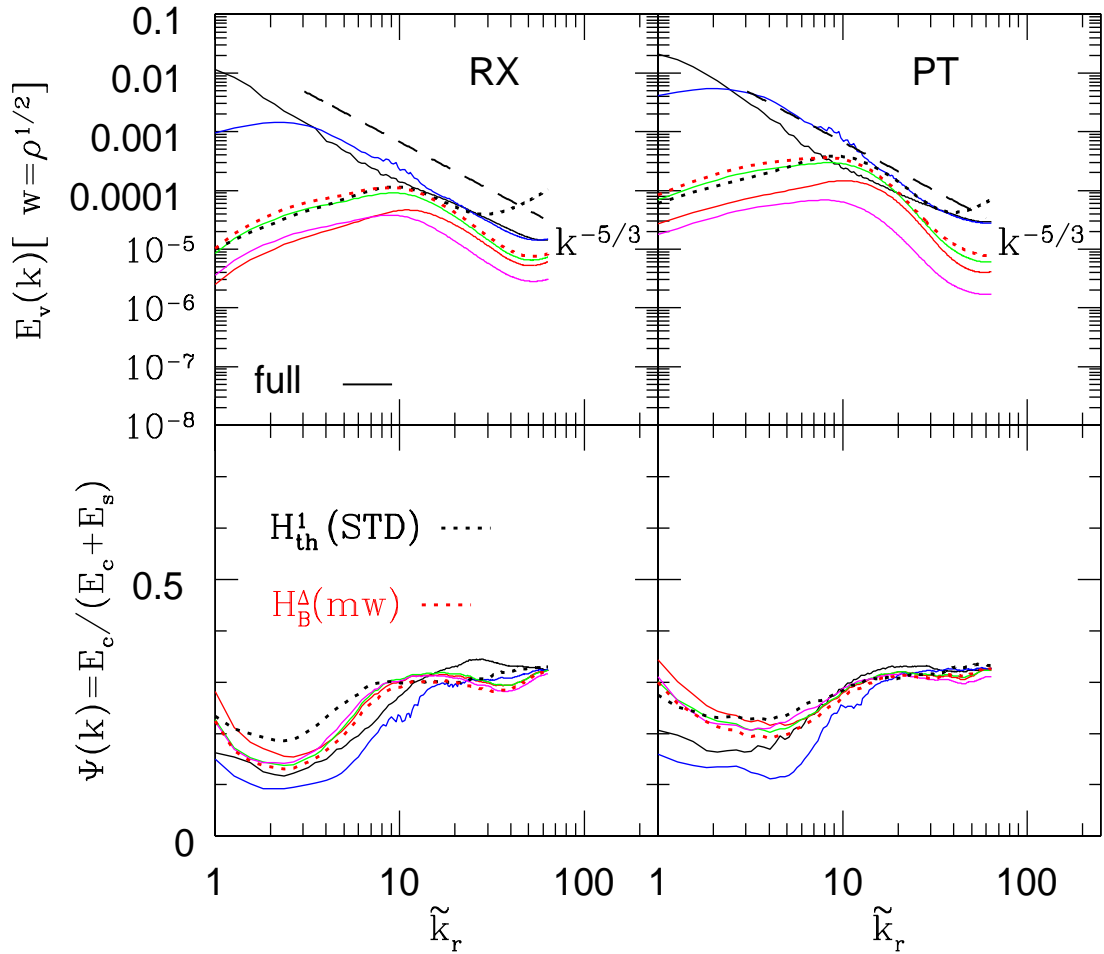}
 \caption{As in Figure \ref{figpwavw.fig}, but for the cooling runs.
}
 \label{figpwcrw.fig}%
  \end{figure*}

Velocity power spectra obtained by applying 
the different filtering procedures  to the simulated cluster 
velocities  are shown in Figures
 \ref{figpwavw.fig}-\ref{figpwcrw.fig}.
Their  behavior  exhibits differences which can be interpreted in terms of 
the variations among the $<H(r)>$   profiles discussed in the previous 
Section. Additionally, we also show (solid black line) the power spectra 
of the unfiltered velocity fields.

The density-weighted spectra of the adiabatic simulations 
(Figure \ref{figpwavw.fig})
are characterized by a peak at ${\tilde k}\sim 10-20$  and a  
power-law behavior $E(k)\propto k^{\alpha}$  
at higher wavenumbers, 
 with a slope $\alpha\simeq -2$, steeper 
than Kolgomorov  scaling ($E(k)\propto k^{-5/3}$).
  These results are in agreement with previous findings 
\citep[][V11]{vaz12}, and indicate how ICM motion becomes turbulent at 
spatial scales $r\simlt 0.1 -0.3r_{200}$. 

The slope $\alpha $  shows some evidence of being  steeper than in the 
unweighted case (Figure \ref{figpwav.fig}). This difference is 
interpreted as being due to the excess power detected at ${\tilde k}\sim 10$
by the density weighting scheme.
The peaks in the power spectra are
common both to relaxed and unrelaxed clusters, with a higher amplitude 
in the PT case due to a greater occurrence of merger  events.

These features of the power spectra of adiabatic runs are shared also 
 by the unweighted spectra of Figure \ref{figpwav.fig}
and can be considered statistically robust, given the size of the subsamples,
suggesting the following scenario.  At cluster scales 
 ${\tilde k}\simlt 1-2$, the ICM motion is dominated by accretion flows from 
large scales, whereas at small scales turbulent motion is driven by
 hydrodynamic instabilities generated by substructure motion and merging events
\citep{ta05,su06}.

The spectral behavior of the filtered spectra exhibits differences 
which are worth investigating in order 
to assess the advantages and shortcomings of the 
 adopted filtering methods. For a constant filtering scale, 
 application of  the filtering procedure (\ref{vfilter.eq})  removes from 
the small-scale velocity field $ \tilde{\vec u}_k$ the spectral components 
 defined by the condition $k H \simlt 1$.  One thus expects the spectral
content of $ \tilde{\vec u}_k$  at small wavenumbers to be further reduced  as 
  $ H \rightarrow 0$.

At large scales the power spectra $\EF(k)$
extracted from the fixed  filtering length set, show a decrease as 
 $ \tilde{k}\simlt 3 $. This is consistent with simple analytical estimates, 
for which the condition $k \HF \simlt 1$ is equivalent to 
 $ \tilde{k}\simlt 10/ 2 \pi $.

Similarly, at large scales the spectra $\EF$ are well above the  spectra of
 all of the other filtering methods which we use.
This is clearly a failure of the fixed length approach,  as can be seen
from Figure  \ref{fighl.fig}: at $r\simlt 0.2 -0.5 r_{200}$ one has either 
 $\HF> H_{th}$ or   $\HF > H_{B}$ which leads the corresponding spectra 
to approach the unfiltered case.
This  was already noticed by \citet{vaz12}, for whom  the agreement between 
the fixed filtering length and multifiltering becomes worse at cluster scales.
This discrepancy is due to the coexistence at large scales of both 
  laminar infall  and chaotic motion, with the fixed filtering 
method missing the velocity correlations.

As one can see from Figure  \ref{fighl.fig}, application of the shock masking 
procedure to the $H^1_{th}$ filtering leads to profiles 
 $H^{(ns)}_{th}(r) $  which are quite close to the unmasked one :
 $H^{(ns)}_{th}(r) \sim  H^1_{th}(r)$ , at least for 
 $r\simlt r_{200}$.  This is at odds with what seen in 
Figure \ref{figpwavw.fig},
where the amplitude of the corresponding power spectra $E_{th}^{ns}(k)$
(solid magenta) is systematically smaller than that of 
  the unmasked case $E_{th}(k)$ (solid green). 
 \begin{figure*}[!ht]
 \centering
\includegraphics[width=17.2cm,height=8.0cm]{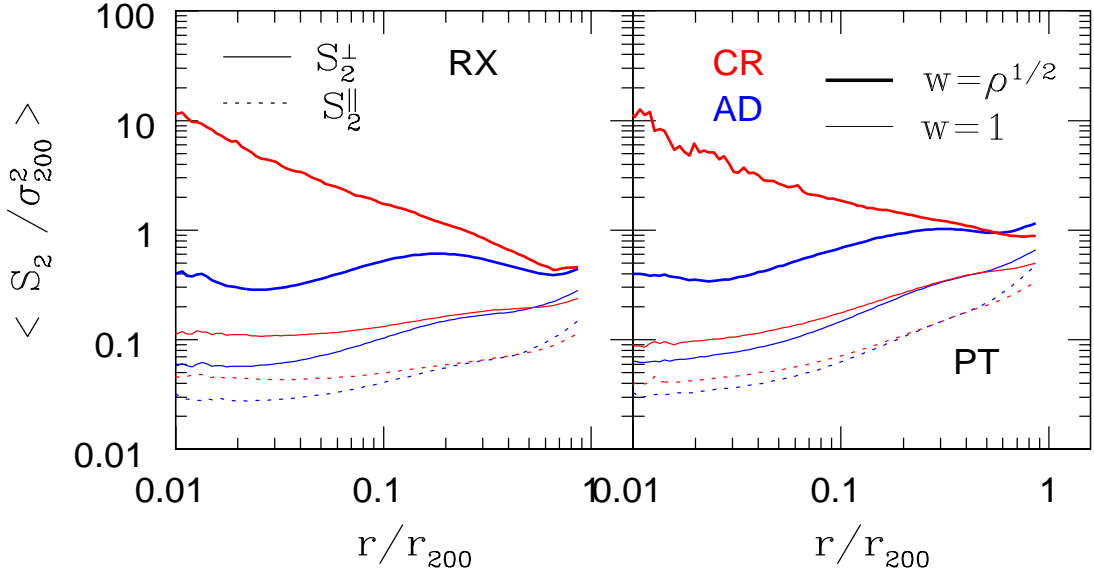}
 \caption{ The ensemble averaged second-order longitudinal and 
transverse velocity structure functions
are shown at $z=0$ as a function of $r/r_{200}$.
The left (right) panel refers to the relaxed (perturbed) subsample.
In each panel the profiles extracted from  adiabatic (cooling) 
runs are shown in blue (red). 
The thick solid lines show the transverse density-weighted velocity structure 
functions. 
}
 \label{figvel.fig}%
   \end{figure*}

This difference in power spectra is due to the weighting scheme used to
evaluate $E(k)$. If the spectra are density-weighted as in 
Figure \ref{figpwavw.fig}, then application of a shock limiter  
preferentially removes from filtering averages the high density particles.
This is confirmed by Figure \ref{figpwav.fig}, where the spectra are 
volume-weighted and the two power spectra 
 $E_{th}^{ns}(k)$ and $E_{th}(k)$ sit on top of each other.

Similarly, differences between spectra $E_{B}(k)$ extracted from 
 $H_{B}$ filtering must be interpreted as being due to a weighting effect.
From Equation (\ref{gauss.eq}) we have seen that evaluation of the filtered 
velocity  (\ref{filt.eq}) is equivalent to a convolution with a Gaussian, 
with similar half-width if the smoothing kernel radii satisfy the 
equalities  (\ref{hsigma.eq}).

From Figure \ref{fighl.fig}, one has $H_{B}(r) \simgt H_{th}(r) $ and 
for the corresponding spectra in Figure \ref{figpwavw.fig} (solid red)
this would imply $E_{B}(k)\simgt E_{th}(k)$.
This is not verified, in fact Figure \ref{figpwavw.fig} shows spectra 
derived from $H_{B} $  filtering that are below those extracted from 
the TH filters. We argue that this behavior can be explained by differences 
in the adopted weighting scheme.

In the TH case, we set the filter function to 
$G(|\vec x_i-\vec x_j|,H_i^n) =m_j $, so that the  velocity
 $\bar {\vec v_{i}}^n $  in Equation (\ref{vfilter.eq}) is a 
mass-weighted average.
In the B-spline filtering procedure 
 $G(|\vec x_i-\vec x_j|,H_i^n) =m_j W_{ij}(H^n_i/\zeta) $, 
which is the SPH density estimate of particle $j$ at point 
 $\vec x_i$. In this case velocity averages are density-weighted 
and the particles $j$ nearest to particle $i$ are weighted more.

One can regard this smoothing procedure as equivalent to a TH 
smoothing but with an effective radius $ < H_{B} $  which, 
accordingly, leads to final spectra  
 $E_{B}(k)$ smaller than in the mass-weighted case.
To verify this conclusion, we constructed a set of filtering lengths 
 $H_{B}(mw)  $   by setting $G=m_j$ and computed the corresponding power 
spectra. These are shown in Figures \ref{figpwavw.fig} to \ref{figpwcrw.fig} 
( red dots) and consistently follow the spectra 
$ E_{th}(k)$.

In order to assess the accuracy of the numerical method which we use, 
we have applied the TH filtering to an ensemble of 
clusters simulated using a standard SPH code. 
The corresponding power spectra are shown in Figures
 \ref{figpwavw.fig} to \ref{figpwcrw.fig} and are indicated 
as  $H_{th}(STD)  $  (black dots).
Their spectral behavior demonstrates that 
the use of a higher-order method,  such as ISPH, 
is crucial at small-scales in order to ensure an accurate modeling of 
turbulence.

The density-weighted spectra of Figure \ref{figpwavw.fig},  
 for standard SPH show excess power at high wavenumbers  which is 
absent in the ISPH runs. 
This power arises from zeroth-order errors  which are intrinsic to 
standard SPH, and in turn impact the modeling of vorticity.
Similarly, at small wavenumbers, the longitudinal-to-total 
 ratio $\Psi(k)\equiv E_c(k)/(E_c(k)+E_s(k))$ 
(Figure \ref{figpwavw.fig}, bottom left) is higher than in the ISPH 
runs. This shows that the problem of properly accounting  for the 
solenoidal part of the spectrum is not a resolution issue.
In a previous paper (V11), it was argued that numerical resolution is 
  critical  when describing the solenoidal part 
  $E_s(k)$ of the velocity power spectrum.
The results presented here  demonstrate that another key role  is played 
by the  numerical method adopted.

{ 
These discussions on  the behavior of power spectra can be considered of 
general 
nature provided that 
the wavenumber dependency of an averaged  spectrum is common to  
the corresponding power spectra of all the subsample clusters.
This in turn implies that at each wavenumber the 
variance of an averaged power spectrum must be sufficient small.
Here, the term sufficient is intended to mean that the area enclosing the 
$1\sigma$ power spectrum dispersion should retain the same spectral behavior 
exhibited by the averaged spectrum.

To confirm the correctness of our conclusions, for the
volume-weighted power spectra $E_{th}(k)$ we then show in the top 
panels of Figure \ref{figpwav.fig} the areas enclosing the 
 one sigma dispersion around the means.
As can be seen from the Figure, 
at each wavenumber the depicted range of power spectrum values is 
relatively small. This justifies the general character of our conclusions 
on the spectral behavior of the considered power spectra.

}

The wavenumber dependency of the ratio $\Psi(k)$ shows that ICM turbulent 
velocities are mostly solenoidal at large scales $ \tilde{k}\simlt 10  $, 
whilst at smaller wavenumbers the compressive component rises to 
 $\Psi(k)\simeq 0.3-0.4$. We interpret this as a genuine feature of the 
measured spectra, and not as being due to a resolution effect (V11).
The bending in $\Psi(k)$ occurs at approximately the same wavenumbers 
 which characterize the maxima of the filtered power spectra. 
This is indicative of how turbulent motion at small scales is sourced by 
substructure motion and merging events, with small-scale shocks raising 
the compressive component of the velocity power spectrum.

Differences among the $\Psi(k)$  referring  to different filtering methods, 
can be interpreted in terms of the differences between the corresponding 
power spectra. In particular, for TH filtering the behavior of $\Psi(k)$  
is in agreement with previous results \citep[see, Figure 8 of ][]{vaz17}.

To summarize, the identification of the correct filtering strategy  
to be applied to the ICM velocity field  depends critically on a 
number of issues. 
Numerically, the starting root $H^0$, as well as  the 
search step $\Delta H$, should be chosen to be 
 as small as possible in order to avoid 
possible biases in the final root values when in the 
presence of velocity fields
with complex patterns. 

Finally, the results presented here show that another critical
feature is in the way in which the velocities in Equation (\ref{vfilter.eq}) 
are weighted, rather than in the choice of the filtering function itself.
This ambiguity is somewhat characteristic  of turbulence and the choice of 
velocity weighting contains a degree of arbitrariness, which depends
on the problem under consideration. As we will see in the next Sections, 
 the top-hat filtering with mass-weighted velocities  ($H^1_{th}$)
seems to produce the most robust and unambiguous results.

In addition to spectral analysis, the second-order velocity structure function 
$\mathcal{S}_2(\vec r)\propto r^{\gamma}$ provides information
in physical space about the small-scale velocity field self-correlation. 
For homogeneous isotropic turbulence one has 
$\mathcal{S}_2(\vec r)\propto r^{\zeta_2}$, with $\zeta_2=2/3$. 
 \begin{figure*}[!ht]
 \centering
\hspace{-1.2cm}
\includegraphics[width=16.2cm,height=8.0cm]{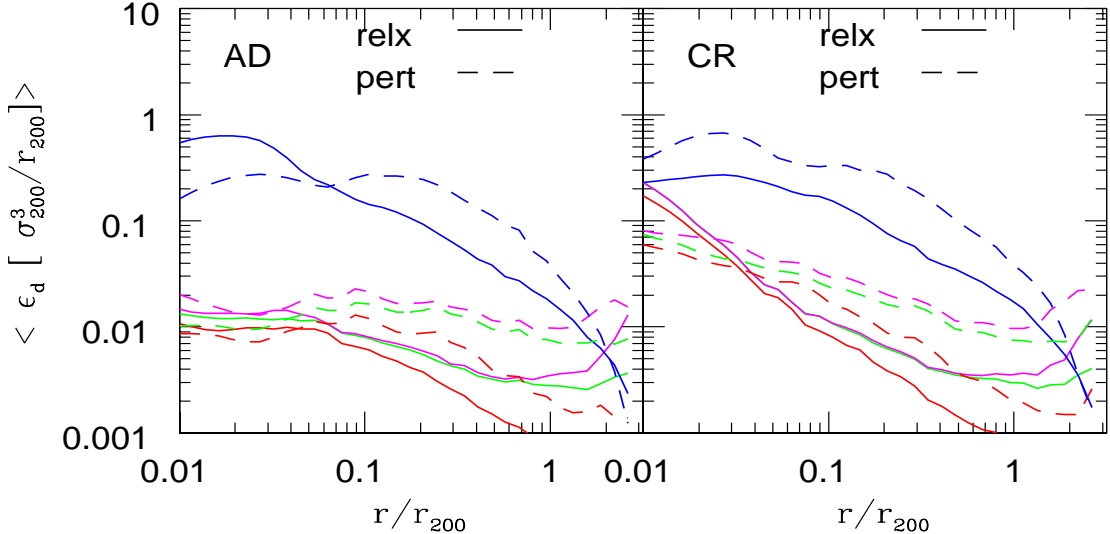}
\vspace{-1cm}
 \caption{
Average radial profiles of the turbulent dissipation rates 
$\varepsilon_d= \delta v^3/l$ , 
are shown at $z=0$ for adiabatic ( radiative) 
simulations  in the left (right) panel. Solid (dashed) lines are for the 
relaxed ( perturbed) subsample. 
The color-coding indicates the adopted filtering as in Figure \ref{fighl.fig} :
$H_B^{\Delta}$  (red), $H_{th}^1$ (green), $\HF$ (blue),
$H_{th}^1(ns)$ (magenta).
The rates are in units of $\sigma^3_{200}/r_{200}$.
}
 \label{figeps.fig}%
  \end{figure*}

For the two cluster subsamples we show separately in Figure  \ref{figvel.fig}
the  parallel and transverse second-order velocity structure functions.
These are computed by using both  volume-weighted and density-weighted 
velocities. All of the volume-weighted functions increase with increasing radii 
following a power-law behavior, with a slope significantly shallower than
$\zeta_2$ for relaxed clusters and approaching $\zeta_2$ if one considers
perturbed clusters. 
Similarly, the amplitude ratio  of  the transverse to longitudinal 
  structure functions 
( $S_2^{\perp}\simeq 2 S_2^{\|}$) is almost constant in radii over
 two decades  and  higher ( $1+\zeta_2/2=4/3$) than  that 
expected in the case of homogeneous isotropic turbulence.

The radial behavior of the structure functions can be compared with 
previous works \citep{mi14,mi15,vaz17}.
There is a general agreement, see for example Figure 7 of \citet{vaz17}, 
but with some differences.
Specifically, for the RX subsample we do not find any indication of 
a steepening at large scales in the longitudinal component.
This is not verified for PT clusters, for which there is a hint for 
such a trend at radii approaching $r_{200}$.
We interpret this as a consequence of the presence of shocks in the 
outskirts of unrelaxed clusters, which are absent  form relaxed ones.

However we stress that making a proper comparison of statistical properties 
is difficult because the 
results presented here refer to sample averages performed over a large 
($\simeq 50$) number of clusters, while in previous papers results were 
extracted by analyzing individual clusters.

Density-weighted structure functions 
 exhibit  a much shallower radial behavior  than 
the volume-weighted functions. This was already noticed (V11) and it is a 
consequence of a selection effect.
By using a density-weighted scheme, most of the contribution to the 
evaluation of the structure functions comes from high-density 
particles, which are located in the inner regions of the cluster.
Because SPH is a Lagrangian code, these are the regions where the 
 bulk of the particles are located.

\subsubsection{Power spectra and velocity structures (radiative simulations)} 
\label{subsec:clpowc}

We now repeat the analysis of the previous Section  by applying the 
filtering methods to cluster velocities extracted from the ensemble of 
radiative simulations.
As outlined in Section \ref{subsec:ac}, the physical modeling of the gas 
 then includes radiative
cooling and star formation, as well as energy and metal feedback from
supernovae \citep{pi08}.

We show  in Figure \ref{figpwcrw.fig} the density-weighted power spectra 
 for the cooling runs.  From a comparison with the corresponding spectra of 
 Figure \ref{figpwavw.fig}  for the adiabatic runs, it emerges 
that the spectral 
behavior of these spectra is characterized by a power excess at small scales.
This feature was already noticed in V11, but the size of the samples allows 
it to be put now  on a more robust footing.

This increase at small scales in the amplitude of  the velocity power
spectra   is common both to relaxed and unrelaxed clusters, so that it can
be assumed to be a  general feature of realistic simulations of galaxy clusters 
which incorporate radiative cooling.

The differences in the spectra 
extracted using  different filtering methods  mirror those 
seen in the adiabatic case and will not be discussed further here.
Similarly, we do not show here the volume-weighted spectra. These
have a spectral behavior similar to the density-weighted ones, but 
with less exacerbated features. For these spectra,
 the slope at ${\tilde k}\simgt 10-20$  is close to    
 ${\alpha}\simlt -2$.

It is interesting to note how, for cooling runs, standard SPH badly 
fails  to properly describe  the velocity power spectra  
at high wavenumbers ($ \tilde{k}\simgt 40 $).
 These are characterized by a much higher 
amplitude ( $\sim 5$) than their ISPH counterparts.
This is a clear shortcoming of the standard method, for which the magnitude
of gradient errors translates  into a difficulty in modeling turbulent motion at 
small scales and, in turn,  produces noisier spectra. These discrepancies 
between the two methods will be further discussed in Sect. 
 \ref{singlecl.sec},
and strengthen the view that the use of ISPH is crucial in order 
to ensure a proper
modeling of turbulence in SPH simulations of galaxy clusters.
 \begin{figure*}[!ht]
 \centering
\hspace{-1.2cm}
\includegraphics[width=16.2cm,height=8.0cm]{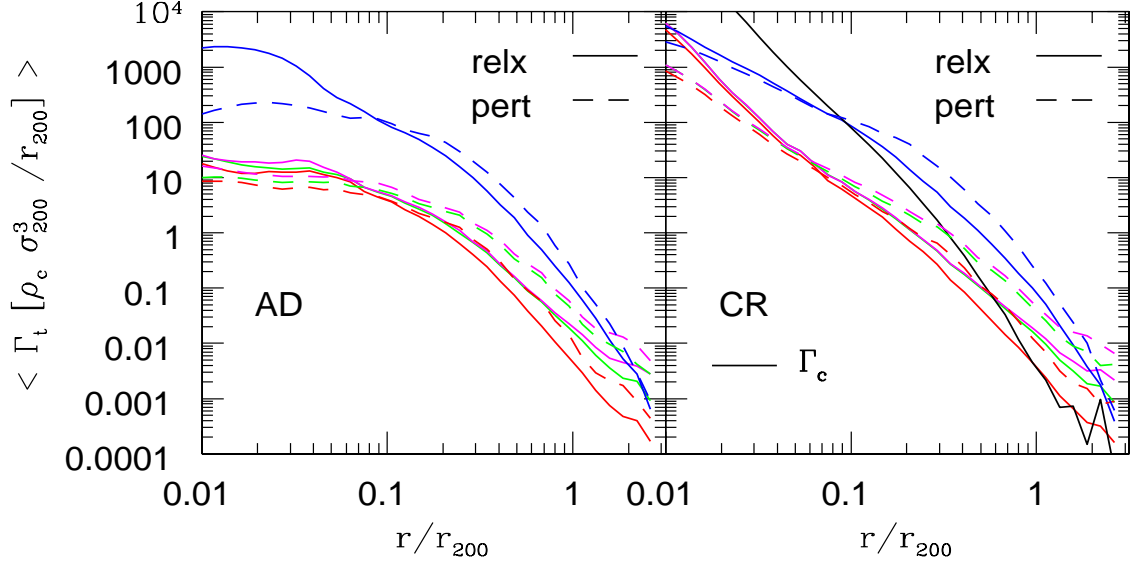}
 \caption{Normalized radial profiles of the turbulent  heating rates
$\Gamma_{t}= \rho_g \delta v^3/l$ are shown 
for the same filtering procedures and simulations as displayed in 
Figure \ref{figeps.fig}. In the right panel, the 
solid black line indicates the average cooling rate profile 
$\Gamma_c=n_e n_I \Lambda(T)$ of the relaxed clusters.
}
 \label{figke.fig}%
  \end{figure*}

We interpret (V11) the power excess seen at small scales 
in the spectra of Figure \ref{figpwcrw.fig}  
  as originating from the development of a dense, compact, 
gas core in the  central region of the cluster.
For cooling clusters, central gas densities are as high as 
 $\rho/\rho_c \sim 10^4$, and are about a factor $\sim 10$ higher 
than in the corresponding adiabatic runs.

Interaction of compact cores with local gas motion triggers 
instabilities \citep{fu04,de05,zu10,ba14}, which in turn generate 
turbulence. That the source of the power excess is due to the presence
of a dense core  is confirmed by the radial behavior of the 
  density-weighted velocity structure functions
$\mathcal{S}_2(\vec r)$ (Figure  \ref{figvel.fig}),
 which for both relaxed and unrelaxed cooling 
clusters display  a radial dependence {\it decreasing} with radius.
Volume-weighted structure functions exhibit a very shallow slope $\gamma$, 
consistent with spectral findings 
\citep[$\gamma \sim -(\alpha+2) $,][]{zu16}

These results present a scenario in which turbulence in galaxy clusters is 
 a multiscale phenomenon. The velocity power spectrum has a peak at 
wavenumbers corresponding to length scales $r_{200}/10 \sim 100 -300 ~kpc$, 
and this is the injection scale which drives turbulence through 
merging and substructure motion. The turbulent motion at 
large scales is mostly solenoidal.  

At small scales there is the second injection mechanism, in which  
gas is stirred   through the interaction of the medium with the core.   
This is the gas sloshing scenario, in which turbulent heating of 
the ICM has been proposed as a viable mechanism to offset 
radiative cooling \citep{fu04,de05,zu10}.
Between these two scales one has subsonic turbulence in 
a compressible medium, but with a power spectrum having a  slope which 
is found  to be close to or steeper than that of Burgers 
turbulence ($\alpha=-2$).

To study the physics of turbulence in  galaxy clusters, several authors 
\citep{yo14,zu16} have constructed mock observations of second order 
structure functions in the presence of multiple energy injection scales.
\citet{yo14} argue that the ability to distinguish the injection scales 
in the projected functions depends critically 
on the relative  heights of the peaks, as well as  on the spatial separation
between the injection scales.

{ 
The construction of mock X-ray maps of gas velocities and related 2D 
structure functions is a non trivial task, which is beyond the 
scope of this paper. Here we just note that  the volume-weighted 
structure functions displayed in Figure  \ref{figvel.fig} can be 
considered as being a realistic expectation of what can be measured from
observations. See, for example, the similarity with the 
projected structure functions
for the two-injection scale model shown in Figure 13b  of 
\citet{zu16}.
}

\subsubsection{Turbulence related profiles } 
\label{subsec:clturb}

We now investigate the radial behavior of some ensemble averaged 
quantities which can be useful metric indicators for turbulence.

For the same filtering procedures previously considered, we first evaluate the 
 turbulent dissipation rates $<\varepsilon_d(r)>$.
Sample averages are computed  by constructing individual cluster profiles 
  $\varepsilon_d(r)\simeq  \delta v^3/l$. These are obtained at each 
test radius by introducing a spherical shell with $40\times40$ grid points 
$\vec x_g$, uniformly spaced in $\cos\theta,\phi$. 
We then compute at each grid point $\vec x_g$ the small-scale
velocity field $\delta \vec v ( \vec x_g)$  and filtering lengths
$ l (\vec x_g)$, estimated from individual particle values 
according to SPH prescriptions. Spherical averaged quantities 
 $\delta  v ( r)$  and $ l (r)$ are then  
defined by averaging over the grid points.
The average radial profiles of other quantities are constructed 
according to the same procedure.

 \begin{figure*}[!ht]
 \centering
\hspace{-1.2cm}
\includegraphics[width=16.2cm,height=8.0cm]{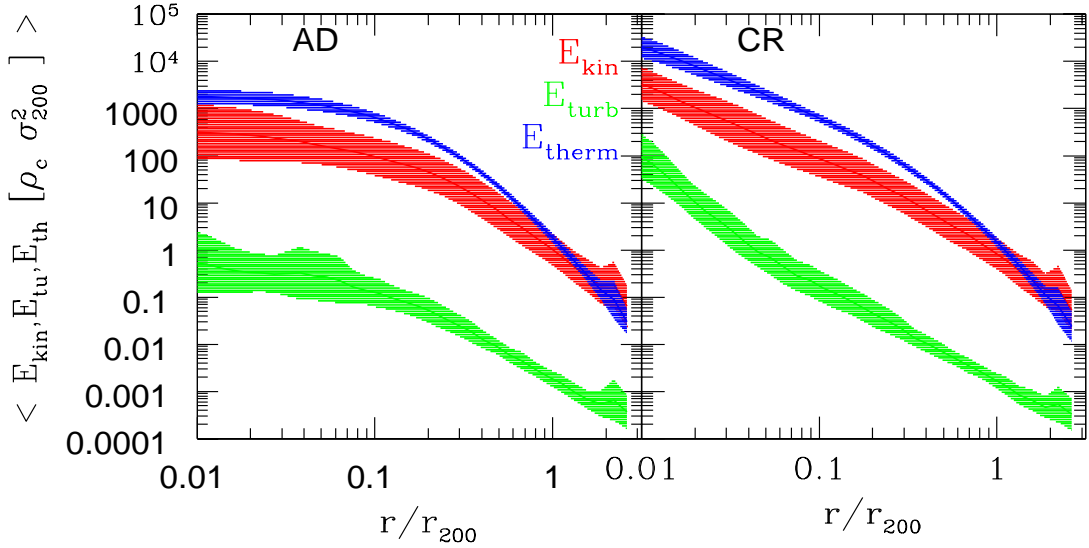}
 \caption{
 Final energy density radial profiles extracted from the relaxed
subsample.The left panel is for adiabatic runs and the 
right panel is for cooling runs.
In each panel are shown: the profile of the thermal energy density 
 $E_{th}= {3 k_B T(r) \rho_g(r)}/{2 \mu m_p} $, the kinetic energy density
 profile $E_{kin}=\rho_g \vec v^2/2 $ and the turbulent one 
$E_{turb}=\rho_g {\delta v}^2/2$; 
the turbulent velocity field refers to  TH filtering.
 The profiles  have been rescaled in units of $\rho_c \sigma_{200}^2$, 
shaded areas represent the $1\sigma$ dispersion range.
}
 \label{figkec.fig}%
  \end{figure*}

The turbulent dissipation rate $\varepsilon_d(r)\simeq  \delta v^3/l$, 
and subsequently the turbulent heating rate 
 $\Gamma_t(r) \simeq \rho_g(r) \varepsilon_d(r)$, 
can be considered as being  robust indicators of turbulence \citep{vaz17}.
In the Kolgomorov scaling regime one has 
$ \delta v \propto l^{1/3} $, so that $\varepsilon_d$ should be
scale independent.
We have seen in the previous Section that estimated power spectra do not  
follow Kolgomorov scaling in the inertial range. 
($ \propto k^{ -5/3}$).  Nevertheless, $\varepsilon_d(r)$  is still a very
useful quantity, since its radial behavior provides spatial 
informations about the energy budget of turbulence, as well as 
about  its deviation from the Kolgomorov regime.

For some of the adopted filtering strategies,  we show the 
corresponding averaged profiles $<\varepsilon_d(r)>$ in 
Figure \ref{figeps.fig}. The left panel is for adiabatic simulations and 
the right panel is for cooling runs.
Within each panel, solid ( dashed) lines are for the $\varepsilon_d(r)$'s 
referring to relaxed (perturbed) subsamples. To consistently perform 
averages between clusters , the cluster values of 
$\varepsilon_d(r)$ have been rescaled 
to dimensionless units:
 $\varepsilon_d \rightarrow \varepsilon_d r_{200}/\sigma_{200}^3$.
A first result to be inferred from Figure \ref{figeps.fig} is that 
the fixed length scale method grossly overestimates the dissipation 
rates.  This is not surprising, given the spectral results already
discussed. 

For adiabatic simulations, there is some hint of Kolgomorov scaling 
only in the case of unrelaxed clusters. For these clusters 
 $<\varepsilon_d>$ stays nearly constant over a radial range of 
about two orders of magnitude.
This is verified only for the $\varepsilon_d$'s extracted from
  $H^1_{th}$  filtering, and with the related shock-limiting procedure 
  $H^1_{th}(ns)$. 
 Over the same range of scales, the condition 
 $\varepsilon_d\simeq const$ is not sustained  by the rates corresponding 
to the $H^{\Delta}_{B}$  procedure.
This is a failure of this filtering method, and illustrates how 
root finding and weighting schemes can introduce biases in the 
final root filtering length values.

For relaxed clusters the condition 
 $\varepsilon_d\simeq const$ holds to a lesser extent, with 
 $\varepsilon_d (H^1_{th})$   dropping from 
 $\varepsilon_d\simeq 8 \cdot 10^{-3} $  at $ r=10^{-2} r_{200}$ 
down to  $\varepsilon_d\simeq 2 \cdot 10^{-3} $  at $ r\simeq r_{200}$.
This deviation from Kolgomorov scaling arises because the 
corresponding power spectra are steeper than in the unrelaxed case. 
We suggest that this is a natural condition for turbulence in a 
steady-state ICM, with the excess power sourced by merging activity 
bringing the spectra to approach the  Kolgomorov scaling.

The dissipative rates $\varepsilon_d [H_{th}(ns)]$,    obtained by 
applying  shock masking to the filtering procedure,  begin to deviate
and become higher than in the unmasked case at $ r\simgt r_{200}$.
This result holds for both relaxed and unrelaxed cases, 
independently of weather one is considering adiabatic or radiative simulations.
We have verified that it is not due to a resolution effect, 
 by running  a high resolution simulation 
 for a  individual cluster  (Sect.  \ref{singlecl.sec}).
 The  dissipative rates extracted
from the simulated clusters were  compared with the corresponding ones 
   from the standard run, obtaining very similar values.

Our results then indicate  that at large radii  $ r\simgt r_{200}$, 
dissipative rates tend to be underestimated if the filtering 
estimator is applied without a shock limiter. At these radii, application of 
a shock limiter is mostly effective, since the presence of supersonic inflows 
due to accretion shocks is significative, and leads
to filtering lengths $H_{ns}$ smaller than $H_{th}$ 
(Sect.  \ref{subsec:clfilters}). From previous results concerning 
  velocity structure functions (Sect.  \ref{subsec:clpowa}), we  have seen that 
$ \delta v \propto l^\beta$, with $\beta>0$ being some value less than unity. 
Therefore, this implies 
 $\varepsilon_d [H_{th}(ns)] > \varepsilon_d [H_{th}]$  as 
$l_{ns} < l_{th}$.

The dissipative rates of the cooling runs are shown in the right panel of 
Figure  \ref{figeps.fig}. In contrast to the profiles extracted from adiabatic 
simulations,  here the profiles exhibit a steady rise when approaching
the cluster center.
This raise is particularly steep in the case of relaxed clusters, and is 
consistent with the findings  of Sect.  \ref{subsec:clpowc}. 
Accordingly, in the  inner regions of cluster cooling runs,
 turbulence is sourced 
by the interaction of the ICM with high density cores,
 the latter being due to radiative cooling and the subsequent star formation.

To assess in a more quantitative way the impact of radiative cooling 
on turbulence, we look at  the turbulent heating rate profiles 
$\Gamma_{t}= \rho_g \delta v^3/l$.   These are constructed in the same 
way as the dissipation rates, we show   in Figure \ref{figke.fig}  
the profiles $\Gamma_{t}(r)$ 
   corresponding   to the  rates of   Figure  \ref{figeps.fig}.

For adiabatic simulations, the profiles $\Gamma_{t}(r)$ tend to approach 
constant values at small radii, when $r\rightarrow 0$. 
This is in contrast with the behavior of the corresponding profiles 
 $\Gamma_{t}(r)$  extracted from the cooling runs.
The profiles exhibit an approximate power-law dependency 
 $\Gamma_{t}(r)\propto r^{-\gamma}$, with $\gamma>0$, spanning almost   
two orders of magnitude in radius, from 
 $r\simeq  0.01 r_{200}$ up to  $r\simeq   r_{200}$. 
The profiles are steeper for relaxed clusters than for unrelaxed ones, 
with $\gamma \simeq 3$ in the former case.

A crucial issue is to determine whether or not dissipation by 
turbulent heating can balance radiative losses in cluster cores.
This possibility has been proposed by a number of authors 
 \citep{fu04,de05,zu10,ba14,zh14a} as a viable mechanism to solve
the cooling flow problem.
For comparative purposes, we have evaluated 
for the simulated clusters of the relaxed subsample,
 the average radial profile of the gas cooling rate:
$\Gamma_c=n_e n_I \Lambda(Z,T)$, where $\Lambda(Z,T)$ is the gas cooling 
function \citep{vo05}. These are evaluated at the radial bin from 
those  of the  individual particle temperatures and 
metallicities: $T_i,~Z_i$.

The results indicate that the cooling rate $\Gamma_c$  is systematically 
higher than the turbulent heating rate  $\Gamma_t$ at all radii
for which $r\simlt 0.5 r_{200}$. Basically, this is a consequence 
of the smallness of the turbulent velocity field $\delta v(l)$ 
in comparison to the other quantities which enter into the 
thermal energy budget of the cluster cores.
This is confirmed by looking at the energy density radial profiles. 
These have been computed for the relaxed subsamples 
of the adiabatic and radiative simulations, the  corresponding averages are 
shown in Figure \ref{figkec.fig}.
 \begin{figure*}[!ht]
 \centering
\hspace{-1.2cm}
\includegraphics[width=16.2cm,height=8.0cm]{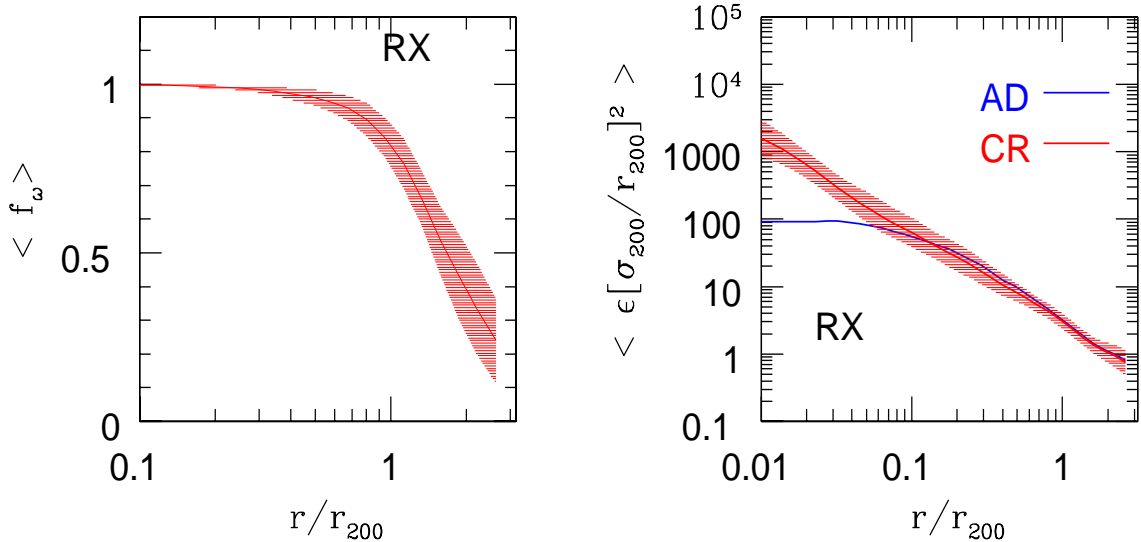}
 \caption{Average final radial profile of the vorticity volume filling factor
$f_{\omega}$  is shown in the right panel for the relaxed subsample of the
cooling simulations. The shaded area represents the limit of the
$1\sigma$ dispersion. The right panel shows the averaged final radial 
profiles of the enstrophy $\epsilon= \omega^2/2$ for the 
relaxed subsamples of the adiabiatic (blue) and cooling (red) simulations.
For clarity, the shaded area representing the $1\sigma$ dispersion 
is shown only for the cooling runs.
}
 \label{figomg.fig}%
 \end{figure*}
 
In each panel are plotted the radial profiles of the 
 the kinetic energy density $E_{kin}=\rho_g \vec v^2/2 $,   
 the thermal energy density $E_{th}= {3 k_B T(r) \rho_g(r)}/{2 \mu m_p} $, 
  and the turbulent  energy density $E_{turb}=\rho_g {\delta v}^2/2$.
 For the thermal energy  $T(r)$ is the mass-weighted
 gas temperature, $k_B$ is the Boltzmann constant and $m_p$ is the proton mass;
 the turbulent velocity field $\delta v$ refers to the TH filtering. 

From the profiles, one sees that the ratio  
 $E_{turb}/E_{th}$  is always $\simlt 1\%$ across all the cluster radius.
For the cooling runs, at  $r\simeq 0.1 r_{200}$ one even has 
 $E_{turb}/E_{th}\simeq 10^{-4}$. This is in contrast with previous findings,
see for example Figure 16 of V11. For the test clusters considered there, 
 $E_{turb}/E_{th}\simeq 2-5\%$. 
This discrepancy is clearly due to  the  use of a 
multifiltering approach; 
in V11  a fixed filtering length  was used for which we have seen 
(Sect.  \ref{subsec:clpowa}) that  turbulent velocities can be significantly 
overestimated.

Given the importance of the topic, we defer  discussion on this
 to the next Section. There, for a single highly-relaxed cluster 
we will study in detail the profiles of some turbulence related quantities.

From previous results on power spectra we have seen that turbulence in the ICM 
is dominated by solenoidal motion, which is characterized by the vorticity
 $\vec \omega= \vec \nabla \times \vec v $. A useful quantity used to quantify 
solenoidal turbulence is the vorticity magnitude, or enstrophy:

\be 
 \epsilon= \frac{1}{2} \omega^2~.
 \label{enstr.eq}
\ee
In accord with previous studies 
\citep{mi14,po15,sch16,vaz17,iap17,wi17}, we will use this quantity to obtain
spatial information about the nature of ICM turbulence.

A complementary measure used to characterize turbulence is the 
volume filling factor. In mesh based codes, 
this is the volume fraction of the cells which  satisfy the conditions 
   $\omega_i > N/t_{age}(z) $, where $N$ is the number of eddy turnovers and 
is set to $N=10$ \citep{mi14,iap17}. At the present epoch the condition
becomes $\omega_i > N H_0$.
In the SPH framework we then define the
volume fraction $f_{\omega}$ as  

\be 
 f_{\omega}=  \frac{\sum_i f_i  V_i} {\sum_i V_i}
 \label{fomg.eq}
\ee

where $f_i=1$ if $\omega_i > N H_0$ and zero otherwise, 
 $ V_i= {m_i}/\rho_i $  and  $\omega_i$ is given by Equation \ref{curlv.eq}.
As for the dissipation rates, we obtain radial profiles  $f_{\omega}(r)$ 
by doing spherical averages of (\ref{fomg.eq}) for the same set of radial bins.

We show  $f_{\omega}(r)$ for the relaxed subsample of the cooling clusters 
in the left panel of Figure \ref{figomg.fig}.  We do not show the corresponding
profile for unrelaxed clusters since its quite similar, but with a 
larger dispersion. The right panel of the Figure for the RX subsample 
shows the radial enstrophy profiles $\epsilon(r)$ 
extracted from  adiabatic 
and radiative simulated clusters. Averages were performed by rescaling 
 cluster enstrophies to dimensionless units : 
 $\epsilon \rightarrow {\tilde \epsilon} =\epsilon (r_{200}/\sigma_{200})^2$.

We can see from Figure \ref{figomg.fig} that the volume filling factor of 
ICM turbulence is very high in the cluster inner regions, with 
  $f_{\omega}(r) \simgt 95 \% $  for $r\simlt 0.5 r_{200}$.  
Beyond this radius $f_{\omega}(r)$ begins to steadily decrease, 
with $f_{\omega}(r)\simeq 80 \% $ $r\simeq  r_{200}$ and 
smaller values at larger radii.

These findings are in qualitative agreement with previous results 
\citep{mi14,iap17}, see for example Figure 8 of \citet{iap17}. 
However, we stress that making a quantitative comparison is difficult 
because  we are presenting here averages extracted from  cluster samples, 
whereas previous papers showed results from a single individual
cluster.

As seen in the right panel of Figure \ref{figomg.fig}, at small radii 
there are 
significant differences between the enstrophy profile $\epsilon(r)$ 
of cooling cluster simulations and the corresponding adiabatic one.
This shows a flat profile for  $r\simlt  0.1 r_{200}$, 
while  for the cooling runs, $\epsilon(r)$ exhibits a 
well defined  power-law behavior  over more than two decades
in radius. This is clearly correlated with the presence of a much denser
core in cooling clusters and is consistent with the findings 
of Sect. \ref{subsec:clpowc}.

For relaxed cooling clusters, the power-law dependency 
of $\epsilon(r)$  is suggestive
of some sort of self-similarity at work. From Kolgomorov scaling one has 
$\omega(l) \propto l^{-2/3}$, but we have seen from the analysis of 
Sect. \ref{subsec:clpowc} that velocity power spectra  are steeper than 
$5/3$.  We thus expect $\omega(l)$ to have a steeper dependency on the 
eddy size $l$. Nonetheless, it is still reasonable to assume a
  dependency of the kind $\omega(l)\propto l^{-\beta}$ with $\beta>0$. 
 This then implies a regulating mechanism for $l$ which sets
the eddy size at radius $r$; we argue that 
 buoyancy forces  due to the strength  of gravity 
can be this mechanism. 
 We will discuss this point in more detail 
in Sect. \ref{singlecl.sec}.
 \begin{figure*}[!ht]
 \centering
\includegraphics[width=17.2cm,height=12.2cm]{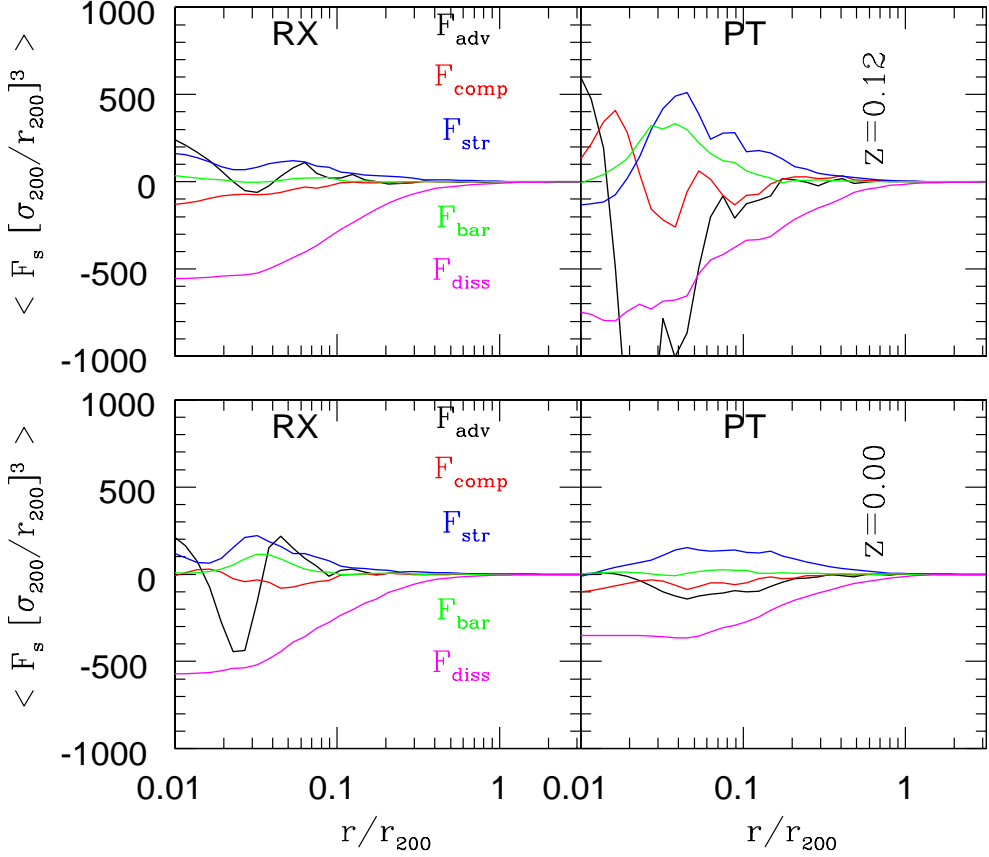}
 \caption{Averaged radial profiles of the different source and sink 
terms present in the enstrophy evolution equation are shown for the 
relaxed ( perturbed) subsample of the adiabatic runs in the left (right) 
 panels. The top panels refer to averages evaluated at $z=0.12$ and 
bottom panels to $z=0$.
}
 \label{figfrsav.fig}%
 \end{figure*}

To investigate the different processes leading to the generation and 
diffusion of solenoidal turbulence, it is also useful to look at the 
 enstrophy time evolution equation, 
this can be written as follows \citep{po15}

\be
\frac{d \epsilon}{ d t}= F_{adv}+F_{stretch}+
F_{comp}+F_{baroc}+F_{diss}~, 
 \label{densth.eq}
 \ee
with the source and sink terms on the rhs of the equation  given by

\be 
\left \{
\begin{aligned}
F_{adv}&=&- \vec \nabla \cdot (\vec u \epsilon)=  
-(\epsilon \cdot \vec \nabla \vec u + \vec u \cdot \vec \nabla \epsilon)~,
\\ 
F_{stretch}&=& \vec \omega \cdot (\vec \omega \cdot \vec \nabla) \vec u =
 2\epsilon ( \vec {\hat \omega} \cdot \vec \nabla )  \vec u  \cdot \vec 
{\hat \omega} ~,\\
F_{comp}&=&-\epsilon \vec \nabla \cdot \vec u = 
- \vec \nabla \cdot ( \vec u \epsilon) + \vec u \cdot 
\vec \nabla \epsilon ~,\\
F_{baroc}&=& \frac{\vec \omega}{\rho^2} \cdot 
(\vec \nabla \rho \times \vec \nabla P )~, \\
F_{diss}&=&\nu  \vec \omega \cdot ( \nabla^2 \vec \omega + \vec \nabla \times 
\vec G)  ~,
\end{aligned}
\label{dfdiss.eq}
\right.
\ee

where $\vec G= \frac{1}{\rho} \vec \nabla \rho \cdot \vec S $, with 
 $ \vec  S$ being the traceless strain tensor \citep{po15}; a hat 
denotes a unit vector and $\nu$ is the numerical kinematic viscosity.
For the latter  we adopt the SPH estimate 
$\nu_i \simeq \alpha_i c_i  h_i /10$ 
\citep{pr12}. The only source term  able to generate vorticity
is  the baroclinic term, aside from dissipative effects, the other 
terms describing the processes of advection, compression and stretching.
SPH estimates of the different terms (\ref{dfdiss.eq}) are computed 
at a given test point $\vec x_g$ from individual particle values.
Radial profiles are then obtained according to the same
procedure as previously described.
 \begin{figure*}[!ht]
 \centering
\includegraphics[width=17.2cm,height=12.2cm]{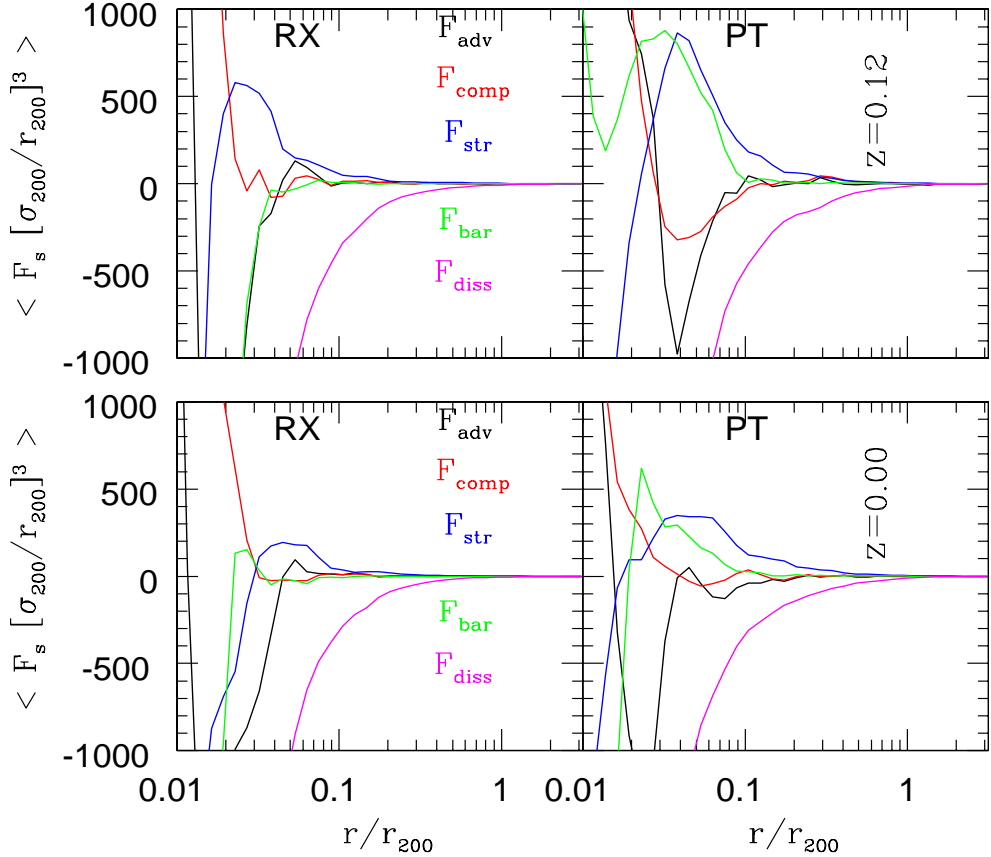}
 \caption{As in Figure \ref{figfrsav.fig}, but for the cooling runs.
}
 \label{figfrscr.fig}%
  \end{figure*}

Previous studies \citep{po15,vaz17,wi17} have investigated, for 
individual clusters, the time 
evolution of the volume averages  of the  source and sink terms present in 
Equation (\ref{densth.eq}). 
 Here we present ensemble average radial profiles 
at two different time slices. 
 Figure \ref{figfrsav.fig}  shows, for adiabatic simulations, the 
mean profiles of the terms (\ref{dfdiss.eq}), obtained by
averaging individual cluster profiles, at the times $z=0.12$ (upper 
panels) and $z=0$ (bottom panels). The left panels refer to profiles
extracted from the RX subsample, the right panels are for the PT subsample.

Profiles extracted from the PT subsample show the impact of the 
baroclinic, stretching and compression terms  as enstrophy generators, 
with the baroclinic term being the primary source.
All of these terms make a non-negligible contribution to enstrophy
production, see for example Figure 11 of \citet{wi17}.
Relaxed clusters are instead characterized, as expected, by smaller 
amplitudes than the PT profiles.

The most important aspects of the impact 
of the driving terms  (\ref{dfdiss.eq})
on enstrophy evolution  
 have  already been investigated in some detail 
 \citep{vaz17,wi17}. 
We focus here on the  differences between 
the profiles  extracted from adiabatic simulations 
and the corresponding profiles computed from radiative runs;
these are depicted in Figure \ref{figfrscr.fig}.

Source term profiles of the cooling runs are characterized by 
higher amplitudes, however the most pronounced differences are seen 
at small radii $r\simlt  0.1 r_{200}$. 
Whilst at $z=0$ adiabatic profiles  of the different terms (\ref{dfdiss.eq})
exhibit the tendency to approach zero or small values 
(with the exception of $F_{diss}$, but see later); in the same
radial range, cooling run profiles tend to increase/decrease to large values.

\begin{figure*}[!ht]
 \centering
\includegraphics[width=17.2cm,height=7.2cm]{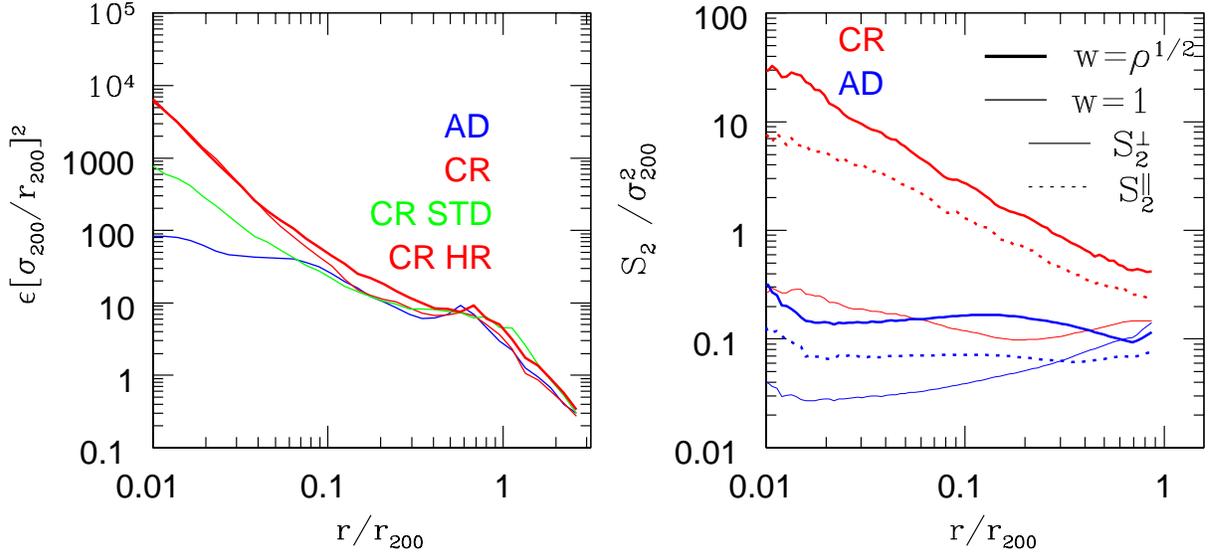}
 \caption{The final enstrophy  profiles of the highly relaxed
cluster with sample index $cl=133$ are shown in the left panel 
for several simulations: adiabatic (AD, blue), cooling (CR, red),
 with cooling but the simulation  run using standard SPH (CR STD, green), 
with cooling but with a simulation performed 
using  an higher resolution  ( $N^{HR}_p\simeq 2N_p$, CR HR magenta).
For the same cluster,  the right panel shows  the radial behavior of the 
 second-order longitudinal and 
transverse velocity structure functions. 
Results from the adiabatic (cooling) run are shown in blue (red). 
The thick lines show the density-weighted velocity structure 
functions. For clarity, in the unweighted case,  only 
 the transverse velocity structure  functions are shown.
}
\label{figomgcl.fig}%
 \end{figure*}

This is clearly a consequence for the cooling runs of the development
of large gas densities in gas cores, which in turn imply a steep rise 
of the compressive term $F_{comp}$ towards the cluster center,  
with the other terms now acting  as sink  terms.
To assess the  robustness of the results against numerical 
resolution, for a single individual cluster we ran a high resolution 
simulation (Sect.  \ref{singlecl.sec}) with an higher number of particles.
We found the final enstrophy profile to be almost identical to that of the 
standard 
resolution run (Figure \ref{figomgcl.fig}), and so conclude that the 
profiles depicted in Figures \ref{figfrsav.fig} and \ref{figfrscr.fig} 
are not adversely affected by resolution effects.

We have verified that the average profile
 $\epsilon(r)$  of relaxed clusters is nearly stationary since $z\simlt0.5$.
This suggests that, at least for relaxed clusters, at late times 
a condition of equilibrium is reached   
with the different terms (\ref{dfdiss.eq}) balancing each other and
  giving   $d \epsilon/dt \simeq0$. 

The very large (negative) values of the dissipative term $F_{diss}$ 
require some explanation.
This term accounts for numerical viscosity effects and in Eulerian
codes it is difficult to estimate its impact \citep{wi17}, 
since there is no  explicit expression for the viscosity $\nu$.
Here, for particle $i$ we estimate  $F_{diss}(i)$  by setting the 
viscosity parameter to $\nu_i \simeq \alpha_i c_i  h_i /10$ 
\citep{pr12}.
This choice  is commonly  used
to estimate the amount of numerical viscosity in SPH simulations
    \citep[see, for example, ][]{pr12b}.

However, it is worth noting that the contribution of particle $j$ to the 
AV tensor (\ref{pvis.eq}) of particle $i$ is applied only when the 
two particles are approaching. This means that by setting
 $\nu_i \simeq \alpha_i c_i  h_i /10$  one puts an upper limit to the 
effective AV viscosity of particle $i$.
We thus conclude that using this estimator in  an equilibrium configuration,
such as that present in cluster inner regions, leads to
a dissipative term $F_{diss}$  which is likely to 
be overestimated.   
The stationarity of the profile $\epsilon(r)$  appears then 
to be driven by the physical terms present in Equation (\ref{densth.eq}).

\subsection{An  individual study of a very relaxed cluster}
\label{singlecl.sec}

To better investigate the impact of turbulence on the properties of 
cool-core clusters, we have decided to look
 carefully at the behavior of some turbulence related quantities
 for a single individual cluster
 extracted from the relaxed subsample of the 
cooling runs, the cluster selection criterion being that of having 
at $z=0$ the lowest subsample value of  $\bar {\Pi}_3(r_{500}) $.
This is  $\bar {\Pi}_3(r_{500}) \simeq -9.4$ and
the cluster has sample index $cl=133$.  
By definition this is also
the most relaxed cluster of the whole  sample. 

The smallness of the  moment $ P_3$  indicates that this cluster, 
for all practical purposes, can be considered perfectly spherical.
We thus expect to gain some insights about the impact of turbulence 
in cluster cores, the cluster being in a well defined highly relaxed state.
Note that it would not have  been possible to identify such a cluster 
without a very large sample size ($ \simeq 200$ clusters). 
For $\Delta=200$, the cluster mass (\ref{mdelta.eq}) is
$ M_{200}\simeq 1.04 \cdot 10^{14}\msun h^{-1}$, so that 
$r_{200} \simeq 1.1 Mpc$  and $ \sigma_{200}\simeq 770 km sec^{-1}$.

We first show in Figure \ref{figomgcl.fig}  the profiles of the 
final enstrophy (left panel) and the
velocity structure function (right panel)  of the chosen cluster.
To check the validity of the simulation results, in the left panel of 
 Figure \ref{figomgcl.fig}  we show the enstrophy profiles extracted from
several additional runs.
In particular, we tested the numerical convergence of the 
 $\epsilon^{cr}(r)$  profile by running a simulation with a higher 
number of particles (HR),  about as twice as  many as in the baseline 
run,  with the other simulation parameters being rescaled accordingly.
We indicate the corresponding enstrophy profile  as 
 $\epsilon^{HR}(r)$.  We also show the profile $\epsilon^{ad}(r)$ 
extracted from the adiabatic simulation, and 
the  $\epsilon^{STD}(r)$  profile of the standard SPH run.

There are several conclusions to be drawn  from the radial behavior of the 
profiles depicted in the left panel of Figure \ref{figomgcl.fig}. 
The first is that $\epsilon^{HR}(r) \simeq \epsilon^{cr}(r)$  throughout 
all of the radial range probed by the plots. This confirms that 
the numerical resolution which we use is adequate to properly describe the 
properties of turbulence  in the ICM of galaxy clusters.

Secondly, the profile of $\epsilon^{STD}(r)$  begins to deviate from 
$\epsilon^{cr}(r)$  at radii $r\simlt  0.1 r_{200}$ and at
 $r\simeq  0.01 r_{200}$ it is smaller than $\epsilon^{cr}(r)$  by almost one order of magnitude.
This is consistent with previous findings (Sect. \ref{subsec:clpowa})
 and happens because the impact of gradient errors is much larger
in standard SPH than in ISPH. This in turn implies, for the 
standard code, a noisier description of vorticity and 
smaller amplitudes of $\omega(r)$ at small scales.
This clearly demonstrates for SPH simulations of galaxy clusters,
the importance  of using ISPH in the modeling of  turbulence.

The adiabatic profile $\epsilon^{ad}(r)$   is depicted for 
comparative purposes, from a comparison with the profiles in the 
right panel of Figure \ref{figomg.fig}, we see that the central value
 of $\epsilon^{ad}(r)$  is close to the ensemble average value ($\simeq 10^2$).
This is in contrast with the cooling run profile $\epsilon^{cr}(r)$, 
 which  has a central value  higher by a factor $\simeq 10$ than
the corresponding ensemble average.
This  suggests that the large value of the central density 
($\rho_g/\rho_c\simeq 10^5$) is closely linked  to the highly 
relaxed state of the cluster. It is also consistent with the proposed
scenario, in which vorticity in the cluster inner regions is driven by the 
interaction of the ICM with the dense compact cores.

Similarly, for the cooling runs, the profiles of the second-order 
velocity structure functions (right panel of Figure \ref{figomgcl.fig}) 
exhibit the same differences with respect the  ensemble average 
profiles of Figure \ref{figvel.fig}.
 \begin{figure*}[!ht]
 \centering
\includegraphics[width=17.2cm,height=7.2cm]{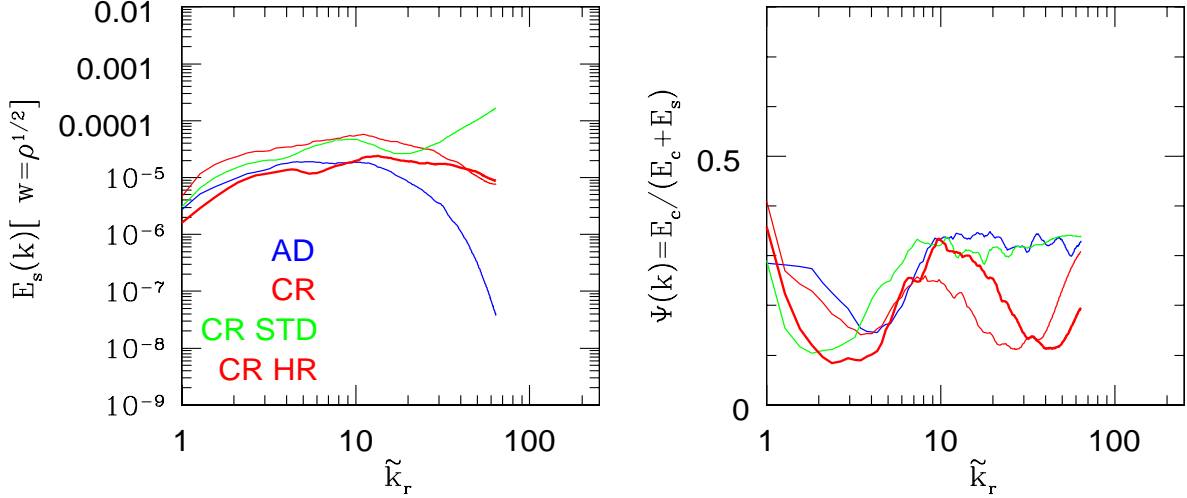}
 \caption{The left panel shows the wavenumber dependency of the
shearing components of the density-weighted velocity power spectra  
for the same cluster simulations 
shown in the left panel of Figure  \ref{figomgcl.fig}.
In the right panel are shown the corresponding 
 ratios of the longitudinal to total velocity power spectra.
}
\label{figpwcl.fig}%
 \end{figure*}

For the same test runs, we show in Figure \ref{figpwcl.fig} the 
density-weighted solenoidal velocity power spectra, together with 
the ratios $\Psi(k)\equiv E_c(k)/(E_c(k)+E_s(k))$. The spectral
behavior of the spectra displayed in the left panel of 
 Figure \ref{figpwcl.fig} is consistent with the findings about
enstrophy profiles. The solenoidal spectra 
$E_s(k)$ and $E^{HR}_s(k)$  are nearly identical across all of the 
 wavenumber range, whilst $E^{STD}_s(k)$ is characterized by 
a power excess at high wavenumbers ($ \tilde{k}\simgt 40 $). 
This is similar to what was already seen in the spectral behavior of 
the ensemble average spectra (Sect. \ref{subsec:clpowc}).

The longitudinal-to-total  ratio $\Psi(k)$ exhibits several features
 which are in contrast with the behavior of the corresponding 
ensemble averaged profile. 
The bottom-left panel of Figure \ref{figpwcrw.fig} shows 
 the ensemble averaged quantity $\Psi(k)\simeq 0.3$ at all the 
wavenumbers  $ \tilde{k}\simgt 10 $.

On the contrary,  from the right panel of Figure \ref{figpwcl.fig} 
we see that  the longitudinal component is 
sub dominant ( $\Psi(k)\simeq 0.1$) in the range 
between $ \tilde{k}\simeq 10 $ and the 
highest wavenumber. This is also valid for the ratio $\Psi(k)$ 
of the high-resolution run. 
We interpret this discrepancy at 
high wavenumbers  along the same line as the difference between 
the central value of the  cluster cooling run enstrophy 
profile $\epsilon^{cr}(r)$ 
and the corresponding sample average value.
For highly relaxed clusters, the large values of central densities act 
as drivers for the generation of vorticity, so that the 
solenoidal component of the power spectrum is dominant.

{
In order to assess the effects of numerical resolution on the 
results presented in this paper, for the test cluster considered here 
 we show in Figures \ref{figomgcl.fig} and \ref{figpwcl.fig} profiles
extracted form a high resolution (HR) run. The simulation was 
performed by running again the ISPH cluster simulation, but with 
a number of particles increased by a factor $\sim$ two. The profiles 
of the HR run are indicated in the Figures with CR HR, 
whereas the profiles corresponding to the standard resolution  run
are labeled as CR.
From Figure \ref{figpwcl.fig} it can be seen that 
at all the wavenumbers   the 
power spectrum extracted from the HR run almost coincides  
 with the  corresponding standard resolution one.

It is also worth noticing  that this result is in stark contrast
with previous findings ( see, e.g., Figure 9 of V11).
In that paper it was argued that to properly describe  velocity power 
spectra  over a decade in wavenumbers,
 at least $N\simgt 256^3$ gas particles are necessary in standard SPH 
simulation of galaxy clusters.

As already noticed in \citet{va16}, this discrepancy follows because
the ISPH scheme is highly effective in removing gradient errors, 
which are dominant at small scales. This in turn implies that ISPH
simulations of subsonic turbulence  exhibit 
at high wavenumbers velocity 
power spectra with a much weaker dependency on numerical resolution.
}

\begin{figure*}[!ht]
 \centering
\includegraphics[width=17.2cm,height=7.2cm]{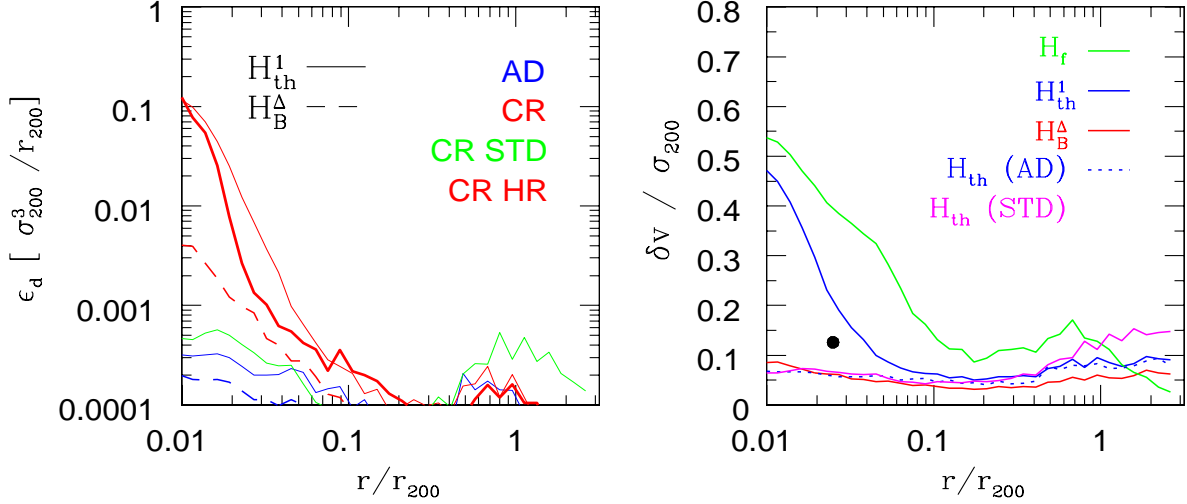}
\caption{ For the same set of cluster simulations depicted in 
 Figure  \ref{figomgcl.fig},  turbulent dissipation rates $\epsilon_d$ 
are shown in the left panel. Different line styles 
indicate different filtering methods:  solid ( dashed) lines refer to
$H_{th}^1$ ($H_B^{\Delta}$).
The radial behavior of different local turbulent velocity fields
$\delta v$ (Equation \ref{dv.eq}) are shown in the right panel as obtained 
by applying different multifiltering procedures to the same 
cluster simulations as in  Figure  \ref{figomgcl.fig}, the color coding 
of the curves is referenced by the corresponding labels.
The black dot   refers to the  {\it Hitomi} velocity measurement
in the Perseus core, rescaled in dimensionless units.
}
\label{figepscl.fig}%
\end{figure*}

Finally the turbulent dissipation rates,  extracted from the suite of 
simulations analyzed here, are shown in the left panel of 
Figure \ref{figepscl.fig}.
We consider profiles obtained by applying to  cluster velocities 
the   $H_{th}$  and $  H^{\Delta}_{B} $  filtering.
  
For TH filtering, the $\varepsilon_d(r) $ profile extracted from 
the high resolution run is very similar to the baseline profile. 
This again confirms the view that the results presented here can be 
considered numerically robust.
Instead, the shortcomings of standard SPH previously discussed now translate 
into  a failure to estimate the dissipation rates at small scales.

At small radii ($r\simlt  0.2 r_{200}$), 
the rates obtained by applying $  H^{\Delta}_{B} $ filtering 
severely underestimate the dissipation rates in cluster cores.
This is a deficiency of the filtering strategy  already discussed 
in Sect. \ref{subsec:clfilters}, and indicates the 
   $H_{th}$    filtering procedure as the optimal choice among the tested
filtering methods.

The turbulent velocity profiles $\delta v(r)$ are shown in 
 units of $\sigma_{200}$ in the right panel of Figure \ref{figepscl.fig}.
The profile $\delta v^{(th)}(r)$  extracted from TH filtering exhibits 
a steep raise as one approaches small radial distances, with a 
central value as high as $\delta v^{(th)}/\sigma_{200}\simeq 0.5$. 
This behavior is shared by the small-scale velocity field 
 $\delta v^{(f)}$ corresponding to the fixed length filtering, but
with a higher amplitude. This is not surprising given the biases 
inherent in this approach. 
  
Both profiles  $\delta v(r)$ obtained either by applying 
 $  H^{\Delta}_{B} $ filtering  or from the standard SPH run, 
are almost flat as a function of radius. This is in line with 
previous findings and the profiles of these runs will not be 
 discussed further.

It is interesting to note that the profile $\delta v(r)$  of the 
adiabatic run, extracted by using TH filtering, is also characterized 
by a very flat behavior. This demonstrates that 
a realistic description of turbulence in cluster cores cannot be obtained
without incorporating at least radiative cooling, and subsequently star 
formation and 
feedback  processes, in the physical modeling of the gas.
In summary, among the profiles depicted in the right panel of 
  Figure \ref{figepscl.fig},  $\delta v^{(th)}(r)$   should represent 
most faithfully the radial behavior of turbulent velocities in a 
relaxed cluster.

To assess whether  the turbulent velocity profile $\delta v^{(th)}(r)$  
is realistic, we can compare it with observations.
To this end we use the first direct detection of gas motion in galaxy clusters 
(H16).
The observations measured, in the core of the Perseus cluster, a 
line-of-sight velocity dispersion of $164 \pm 10~ km s^{-1} $ in the region 
$30-60~kpc$ , and a gradient of $150\pm 70~ km s^{-1}$ 
across the $60~kpc$ central region.

The Perseus cluster is a nearby ($ z\simeq0.0179$), relaxed massive object.
\citet{si11} report  a cluster mass at $\Delta=200$ of 
$ M_{200}\simeq 7 \cdot 10^{14}\msun $,  with 
$r_{200} \simeq 1.8 Mpc$. This implies   $ \sigma_{200}\simeq 1300 km sec^{-1}$.
By rescaling the measured velocities one thus obtains 
 $\delta v/\sigma_{200} \simeq 0.126$  
at the midpoint radius $r/r_{200}\simeq 0.025$.

This value is smaller than the simulation value of $\delta v^{(th)}$   at the 
same radius:  $\delta v^{(th)}/\sigma_{200} \simeq 0.2$, but is not 
grossly inconsistent with it, given the uncertainties involved in the 
comparison.
In particular,  our sample cluster is very relaxed but less massive   
($ M_{200}\simeq 1.04 \cdot 10^{14}\msun h^{-1}$) than Perseus.
Our findings suggest that the presence of turbulence in cluster cores 
is closely linked to the impact of cooling and to the development of 
large core densities. Therefore, at a given radius one expects  to measure
  a relative increase in the strength of turbulence as the cluster 
mass decreases.

Another quantity to be compared with simulation results  is the 
ratio  of  kinetic to thermal energy density in the core.
From measured velocities (H16) this ratio  is found  to lie  in 
the range $\simeq 5-10\%$.
This value  is not in contrast with that obtained from 
 the energy density profiles depicted in the right panel of 
 Figure \ref{figkec.fig}. At $r/r_{200}\simeq 0.05$ the ratio 
is estimated to be $E_{kin}/E_{th}\simeq 10\%$, broadly consistent with 
observations.

However, a critical issue to be kept in mind is that the simulations 
presented here  are purely hydrodynamical, and do not include several
physical processes which can impact on the modeling of turbulence in cluster
 cores. In particular, we do not incorporate AGN feedback.
This is a severe limitation in the case of Perseus, because the cluster is
observed to host a powerful AGN activity \citep{fa11} which is
expected to drive turbulence in the cluster core \citep{zh16}.
 \begin{figure*}[!ht]
 \centering
\includegraphics[width=17.2cm,height=7.2cm]{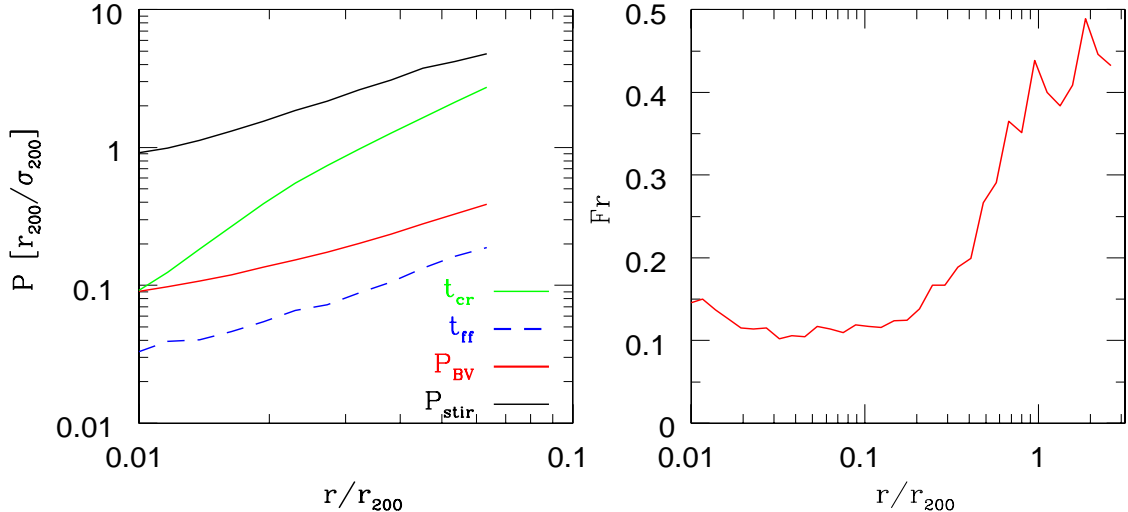}
 \caption{Left: For the cooling run of the highly relaxed
cluster  $cl=133$, several timescales are shown 
as a function of radius. 
The sloshing oscillation period $P_{BV}= 2 \pi / \omega_{BV}$, 
where $\omega_{BV}$ is the Brunt-V\"{a}is\"{a}l\"{a} frequency, 
(solid red line); the
stirring period $P_{stir}= 2 \pi / \omega_{stir}$, 
where $\omega_{stir}= \delta v /l$ , (solid black line);
the cooling time $t_{cr}\simeq (3/2) n k_B T /Q_R$
(solid green line); the free -fall time 
$t_{ff}\simeq 1/\sqrt{\rho}$ 
(dashed blue line). 
Right: the radial behavior of the Froude number 
$Fr\simeq \delta v / (\omega_{BV} l)$.
}
\label{figfrqcl.fig}%
 \end{figure*}

Recently, a number of authors \citep{hi17,la17,bo17,zu18} have investigated 
 the consistency of the low level of gas motion 
observed in the Perseus core with the presence of an 
on-going AGN activity.  
Simulation results have reached conflicting conclusions, in particular 
\citet{zu18} argue that gas sloshing alone is sufficient to reach the 
observed level of gas velocities. We will return on this  topic
in the Conclusions.

We now investigate the radial behavior of some characteristic timescales 
which regulate gas motion in the cluster center.  In a medium at equilibrium
having  ${d ln S}/{d ln r}>0$, a fluid element displaced at the radius
 $r$ from its equilibrium 
position will be driven back by a restoring force and will oscillate  at the 
 the buoyancy or Brunt-V\"{a}is\"{a}l\"{a} frequency  \citep{cox80} 

\be
\omega_{BV}=\Omega_K \sqrt { \frac{1}{\gamma} \frac{d ln S}{d ln r}  }~,
\label{brv.eq}
\ee

where  $\Omega_K= \sqrt{GM/r^3} $ is the Keplerian frequency.

The ICM of a relaxed cluster exhibits density variations by more than 
three orders of magnitude across the whole cluster, and in a 
strongly stratified stable medium the impact of buoyancy forces 
on turbulence is significant. This hydrodynamical regime is well known in 
geophysical fluid mechanics \citep{ri00} and is commonly referred to 
as stratified turbulence.


To quantify the impact of gravity on turbulence it is convenient to 
introduce the Froude number  \citep{ri00}

\be
Fr \simeq \frac{ \delta v }{\omega_{BV} l}~,
\label{frn.eq}
\ee

which measures the  relative importance of buoyancy forces in
comparison to stirring motion. If $\fr \simgt 1$  inertial 
forces are dominant, while when $\fr \simlt1$ 
 the effect of stratification is significant.

In this regime the gas motion is preferentially tangential, being 
suppressed along the radial direction by buoyancy restoring forces.
This follows because the stirring frequency is lower than 
 the  Brunt-V\"{a}is\"{a}l\"{a} frequency :
$ \delta v /l \simlt \omega_{BV} $ and gas motion excites gravity
waves which are trapped inside  the radius given by the condition
${\it Fr} \simeq 1$ \citep{ru10,zh14b}.
The nature of these gravity waves, or g-modes, is found to be 
preferentially tangential \citep{cox80,lu95,ri00}, and the 
regime $\fr \simlt 1$ is then characterized by anisotropic turbulence.

We have evaluated the radial profile of the gas sloshing 
period $P_{BV}(r)= 2 \pi / \omega_{BV} $ 
up to the  maximum radius $r_{max}(S) \simeq 0.08 r_{200}$ 
for which ${d ln S}/{d ln r}>0$. Note that, for the clusters of the 
relaxed subsample, the test cluster $cl=133$  is the one with 
the maximum value of $r_{max}(S)$. To construct the radial profile 
we have estimated $\alpha_S={d ln S}/{d ln r}$ by setting
$\alpha_S=1.1$.

Additionally, we also evaluate the  stirring motion timescale 
 $P_{stir}= 2 \pi / \omega_{stir}\simeq 2 \pi l/ \delta v$, 
the cooling time scale $t_{cr}\simeq (3/2) n k_B T /Q_R$, and the
the  free -fall time $t_{ff}\simeq 1/\sqrt{\rho}$. 
All of these profiles are shown in the left panel of 
Figure \ref{figfrqcl.fig}; the Froude number 
$\fr(r) = P_{BV}(r)/P_{stir}(r)$ 
 is depicted in the right panel.

A clear result which emerges from the plots is that in the probed radial 
range 

\be
t_{ff} < P_{BV} \simlt t_{cr} <  P_{stir}~,
\label{tlim.eq}
\ee

and  there  is a very weak stirring motion with 
a strongly stratified turbulence having $\fr \simlt 0.1$ up to 
 $r \simeq 0.1 r_{200}$.

This result is particularly important since it indicates that, in
the  inner regions of relaxed clusters, turbulence 
is dominated by buoyancy forces  and regulated by the 
Brunt-V\"{a}is\"{a}l\"{a} frequency. Recently   \citet{sh18},
from an adiabatic simulation of a single cluster, obtained similar 
values for the Froude number in the cluster inner high density region.
Turbulent velocities were extracted by applying a wavelet analysis, 
so that their result provides an independent  confirmation of our 
values for the Froude number.

The profiles of Figure \ref{figfrqcl.fig} can be compared with recent 
measurements of the 
Fornax cluster \citep{su17b}. Fornax  is a very close by
$z\simeq 0.00475$ low mass cool core cluster, with approximately 
$r_{200} \simeq 750 ~kpc$ \citep{su17a}.
{\it Chandra} observations reveal that the cluster 
morphology is typical of a sloshing cluster, with  discontinuities 
in the X-ray surface brightness and multiple sloshing cold fronts 
\citep{su17a}.

From a combined {\it Chandra} and {\it XMM-Newton} analysis \citet{su17b} 
derive ICM density and temperature profiles. Under the assumption 
of hydrostatic equilibrium, these profiles were used to calculate 
 the Brunt-V\"{a}is\"{a}l\"{a} frequency  and the sloshing 
period $P_{BV}$ at the radius $r$.
 In addition, they derived also the free-fall and cooling 
time scales. The radial behavior of these profiles in the cluster core 
are shown in their Figure 8, along the x-axis the radial distance 
$R=10~kpc$ corresponds to $r / r_{200}\simeq 0.01$.

A comparison with  the corresponding profiles depicted here in the 
left panel of Figure \ref{figfrqcl.fig} shows a remarkable agreement.
 The ratio  $ P_{BV}/t_{ff}$ is of the order of 
   $ P_{BV}/t_{ff}\simeq 5 $   up to $r / r_{200}\simeq 0.04$.

From their Figure 8 for the dimensionless period 
    at $R=10~kpc$   we obtain    
$ P_{BV}/t_{200}\simeq 2 \cdot 10^2 /1.3 \cdot 10^3 \simeq 0.15$, 
where  we have estimated for the Fornax cluster 
$t_{200} =r_{200}/\sigma_{200} \simeq 1.3 \cdot 10^3 Myr$.
This is close to the value of $ P_{BV}/t_{200} \simeq 0.1 $ found here at  
$r / r_{200}\simeq 0.01$

The ratio $ t_{cr}/P_{BV}$  is found here to decrease from the range 
 $ t_{cr}/P_{BV}\simeq 5-10 $  above $r / r_{200}\simgt 0.02$ down
to unity at $r / r_{200}= 0.01$. This is in contrast with the 
measured ratio, for which $ t_{cr}/P_{BV}\simeq 5-10 $. 
This divergent behavior at small distances is not unexpected, given 
the absence in the simulations of AGN heating.  

From the X-ray analysis of the Fornax cluster, \citet{su17b} argue that 
gas sloshing contributes to the heating of the core. Their conclusions
are based on the smallness of the measured sloshing timescale
 with respect $t_{cr}$.  The agreement between the timescale radial 
profiles extracted from our simulations, and the corresponding 
profiles shown in their Figure 8 supports this view.
However, it is worth noting  that over the probed radial range, 
the profiles of Figure \ref{figfrqcl.fig} never violate 
the condition $t_{stir} >  t_{cr}$. This implies a condition of
negligible turbulent mixing and , in agreement with the 
findings of Sect.  \ref{subsec:clpowc}, that the contribution of 
sloshing motions to core heating is sub-dominant.

In the sloshing scenario, core heating proceeds via entropy mixing 
driven by the presence of KHI.
The latter is identified in X-ray maps by the presence of KH eddies 
at the cold front interfaces \citep{ma07}.
 Since the  growth of KHI is suppressed in the
presence of a viscosity  or an aligned magnetic field \citep{zu11,ro13}, 
 observational support for the presence of KH eddies \citep{su17b,su17a}
can then be used to put constraints on the ICM physical properties.
The derived upper limits on the ICM Spitzer  viscosity \citep{su17b}
favor an almost inviscid, weakly magnetized, ICM. 
This adds credibility to the  use of the hydrodynamical simulations 
presented here in the   modeling of  ICM turbulence.

\section{Summary and Conclusions}
\label{concl.sec}

In this paper we have presented results concerning
 the properties of turbulent
motions in the ICM. These have been identified by the application 
to cluster velocities of different multifiltering strategies, so as to
separate bulk flows from small-scale motions. 
We have then applied these filtering procedures to gas velocities
extracted from  large sets of simulated clusters. 
The velocity power spectra and radial profiles of turbulence related quantities
have then been contrasted to identify the optimal filtering strategy 
and to provide physical insights into the role of turbulence in the ICM. 
The application of different filtering methods to the same gas velocity sets,
demonstrate that the root filtering lengths depends critically on a
number of issues.

Numerically, it has been found that the values of the final root lengths
 $H_{i}^n$ are quite sensitive to the chosen initial values 
 $H_i^{0} $ as well as to the step lengths $ \Delta H $.
This follows because the algorithm terminates the root search when 
Equation \ref{vtol.eq} is satisfied, thus the root values can be biased to 
high  values
if the interval bracketing  the root is too large.
This occurs in the presence of velocity fields with complex patterns, having
 short range correlations.

Similarly, the final root values $H_{i}^n$ are also determined by the choice
of the filtering function $ G(|\vec x|,H )$.
We have adopted different smoothing functions to construct different sets 
of filtered velocities. The results of Sect. 
\ref{subsec:clfilters} and \ref{subsec:clpowa} demonstrate that 
differences in the corresponding velocity power spectra can be 
consistently interpreted in terms  of the adopted 
smoothing procedures.

In particular, in comparison to TH filtering, B-spline smoothing 
 tends to give more weight to close by particles. 
 This in turn implies higher values
 for the root filtering lengths and lower amplitudes for the 
velocity power spectra. From the results presented in 
Sect.s \ref{subsec:clpowa}  to \ref{singlecl.sec}, we argue that TH
filtering represents, with the search parameters used here, the 
optimal choice to extract turbulent velocities in a variety 
of cluster dynamical states.
In addition, note also that the fixed  filtering length approach clearly
 fails in the presence of complex velocity flows.

Results extracted from  subsamples of the adiabatic cluster simulation
ensemble, delineate the following scenario for the generation and  
evolution of turbulence in the ICM. Power spectra 
of the small-scale filtered velocities exhibit a maximum at 
 ${\tilde k}\sim 10-20$  and a  power-law behavior $E(k)\propto k^{\alpha}$  
at higher wavenumbers. The wavenumbers  ${\tilde k}\sim 10-20$  
 correspond to the injection  scales $r_{200}/10 \sim 100 -300 ~kpc$, 
with turbulence being driven   by merging and substructure motion. 

The slope ${\alpha}$ lies in the range  ${\alpha}\simeq [-3,-2]$, with 
density weighted spectra being shallower than volume weighted ones. 
The turbulent motion
at large scales is mostly solenoidal, with the fraction of longitudinal spectrum 
rising to $\simeq 0.3$ at small scales. Perturbed clusters are characterized 
by a power excess at small scales due to a higher  merging rate. In fact, 
the slopes of their velocity power spectra  are closer to Kolgomorov 
($\alpha \simeq -5/3$), and the estimated dissipation rate 
$\varepsilon_d \simeq \delta v^3/l$  is nearly constant over the whole 
 cluster radial range  tested by the simulations ($\sim 0.01 r_{200}$ to 
$\sim  2 r_{200}$).

In this scenario the generation of vorticity proceeds initially through 
the baroclinic term, and is  sustained by the compression and stretching
terms present in the dissipative enstrophy Equation \ref{dfdiss.eq}.
These results are essentially in agreement with previous  studies
\citep{vaz12,vaz17,iap17,wi17}, the only substantial difference being the 
robustness of our findings   given the size of the samples which we use.

For the cooling runs, the results of Sect.  \ref{subsec:clpowc}
to \ref{singlecl.sec} demonstrate that ICM turbulence properties undergo
a drastic change  with respect those found in adiabatic simulations.
This is due to the development of dense compact gas cores, 
as a consequence of radiative losses. These cores interact with the local 
gas motions triggering instability,  and  in turn leading to the 
development of turbulence \citep{fu04,zu10,ba14}.

When contrasted with adiabatic spectra, velocity 
power spectra of radiative runs are characterized by a much flatter 
wavenumber dependency. We interpret this as being due a consequence of the 
injection of turbulence at two different scales. The first  injection is 
at scales  $r_{200}/10 \sim 100 -300 ~kpc$, which drives turbulence through the 
usual mechanism. The second occurs at small scales and is due to the stirring 
of gas motions through  interaction of the ambient medium with the core.
This spectral behavior was already noticed in a previous paper (V11), 
but it is now statistically significant.  The impact of these physical 
effects is important in several aspects.

The estimated dissipation rates $\varepsilon_d $  increase as $r$ decreases, 
with a very steep radial dependency in comparison to the adiabatic rates.
 In fact, the central values of $\varepsilon_d $  are  higher by almost an 
order of magnitude.
Nonetheless, the corresponding turbulent heating rates 
$\Gamma_{t}(r)= \rho_g \varepsilon_d $  are found to be unable to balance
the cooling rates $\Gamma_c(r) $ in the cluster inner parts 
 ($r \simlt 0.5 r_{200} $ ). This calls into question the viability 
of the turbulent heating model \citep{zh14a} for solving the cooling flow 
problem.
This has  already been noticed \citep{ba14}, and is 
a consequence of  the smallness of turbulent velocities in cluster cores.

Another striking feature that emerges from the analyses of the cooling runs 
 is the behavior of the average radial enstrophy profile 
 $\epsilon(r)= \omega^2/2$. For relaxed clusters the profile is stationary   
since $ z \simeq 0.5$. and exhibits a power law dependency 
 $\epsilon(r)\propto r^{-\beta} $, with $\beta \simeq 3/2$, over more than
two decades in radius. Analysis of the enstrophy dissipation equation
shows that the only source of enstrophy is the compression term,
which is dominant as $r \rightarrow 0$  because of the large gas 
densities now present in cluster cores.

The power-law behavior of  $\epsilon(r)$ implies a dependency of
the vorticity amplitude on cluster radius $r$, which  suggests 
that the eddy size $l$ also depends on radius: $l=l(r)$.
To better clarify this issue we have investigated in detail the turbulence
properties of a single, highly relaxed, cluster. In fact, the cluster 
has been chosen as being the most relaxed of the $\simeq 200$ sample 
clusters.

The cluster has sample index $cl=133$ and from the profiles displayed in
Figure \ref{figepscl.fig} we see that the turbulent dissipation rate and 
velocity profile have a very steep decline with radius.
We have compared the turbulent velocity profile with the only 
available  direct detection of gas motion in a galaxy cluster (H16).
By rescaling the velocities  measured in the core of the Perseus cluster 
we obtain at the estimated radius a dimensionless velocity which is 
$\sim 50\%$ smaller than the simulation value.

We do not consider this discrepancy as problematic, given the mass 
difference between Perseus and the simulated cluster as well as 
 the non-linearities
induced by cooling effects. However, the comparison
 confirms the low level of turbulent motions which we obtain in cluster cores.
Note that, as previously outlined,  it is the smoothing given by TH filtering
which produces the velocity profile in best agreement with observational data.

We have also constructed the sloshing timescale profile and contrasted it 
with recent measurements presented for the Fornax cluster \citep{su17b}. 
In the probed radial range, from $0.01$ up to $0.05 r_{200}$, 
we find  that the sloshing oscillation period $P_{BV}(r)$ and other timescale 
profiles are in very good agreement with the measured timescales.
Additionally, we used the profile $P_{BV}(r)$ to evaluate the Froude
number, which is found to be 
$Fr\simeq \delta v / (\omega_{BV} l) \simeq 0.1 $ up to $r \simeq 0.1r_{200}$.

We use this result, and the agreement  with the Fornax cluster profiles,  to
draw several conclusions about the properties of turbulence in the cores
of relaxed clusters. The smallness of the Froude number indicates that turbulent 
motion in cluster cores is dominated by gravitational 
 buoyancy forces, with stirring motion strongly suppressed along the 
radial direction. In this regime of stratified turbulence, 2D stirring
motions are very weak, and the energy injected is unable to sustain the 
cooling rate.
We consider these findings as the main result of our paper.

The simulations presented here have a number of limitations which we
discuss now.  We have contrasted power spectra and turbulence related 
profiles against the corresponding ones extracted from clusters simulated 
using standard SPH. The results demonstrate the superiority of ISPH 
in the numerical modeling of subsonic turbulence. This is because 
it is crucial  to keep gradient errors very small in order to 
ensure a correct description of vorticity \citep{va16}. 

In particular,
 a previous paper (V11)  argued for a strong dependence of 
solenoidal velocity spectra on numerical resolution and particle number.
For the test cluster $cl=133$ we found numerical convergence 
(Figure \ref{figpwcl.fig}) between 
the solenoidal velocity power spectra extracted from the baseline sample, 
and a high resolution run with about twice the number of particles. 
 This demonstrates that the numerical resolution used here is adequate
to describe velocity power spectra over more than a decade in wavenumber.

Whilst the numerical scheme which we use can be considered adequate to 
simulate ICM hydrodynamics, our physical modeling of the ICM is incomplete.
Our baryonic physics incorporates radiative cooling, star formation  and
supernova  feedback, but assumes an inviscid, unmagnetized, ICM. 
We  consider  these assumptions realistic,  since X-ray observations of 
KHI \citep{su17a,su17b} favor a low viscosity plasma with low values 
 of ordered magnetic fields.

However, we do not include ICM heating due to AGN feedback.
This model  is widely  supported both observationally and theoretically  
\citep[see, for example, ][and references cited therein]{zw18}.
In this scenario the  ICM is thermalized 
 by means of the interaction with buoyantly rising bubbles created from 
AGN jet activities. 
A variety of mechanisms have been proposed to transfer the
 feedback energy  to the ICM \citep{zw18}, but it is still  
unclear which are the dominant processes driving the heat transport.
In particular,   \citet{ya16b} argued, on the basis of purely hydrodynamical
simulations, that turbulent heating from AGN feedback is negligible.

We conclude that our simulation results, which are consistent with 
available data,
seem to rule out the turbulent heating model as 
 the sole   solution for the 
cooling flow problem. 
Similar conclusions have been recently reached by \citet{mo18}.
From a set of homogenous isotropic turbulence simulations, which
incorporate radiative cooling, the authors conclude that
consistency with {\it Hitomi} results rules out turbulent heating models
as the dominant source of heating in cool core clusters.

It then appears that the solution to this problem
 is highly non trivial,  and will require galaxy cluster simulations of 
increasing physical complexity. In a Lagrangian framework, the adoption of 
ISPH 
is undoubtedly  a step in the right direction.


\appendix
\section{The shock detection algorithm}

Following \citet{be16b}, shocks in SPH simulations are  found 
as follows.  When a shock front develops the local velocity exceeds the sound 
velocity, and the shock front separates the pre-shock (upstream) from the 
post-shock (downstream) regimes. The implementation of an SPH shock-finder
 requires first the identification of the shock direction and then 
the evaluation 
for particle $i$ of the hydrodynamic variables in the two regimes. 

We identify for particle $i$   the direction of the
shock normal  $\vec n$ as given by the pressure gradient: 
$\vec n_i =-\vec \nabla P_i/P_i$, where

   \be
 \vec \nabla P_i =   \frac{1}{\rho_i} \sum_j  m_j (P_j-P_i) 
\vec \nabla W_{ij} 
   \label{grpres.eq}
   \ee

is the SPH estimator for the pressure gradient \citep{pr12}. 
The normal $\vec n_i$ points
in the downstream direction and  the distances of the two 
regimes from particle $i$ are then 
$\vec x^{d,u} _i= \vec x_i \pm  \zeta h_i \vec n_{i}$, 
where one is assuming that the shock extends throughout the 
kernel domain. Setting   $\vec x_i-\vec x_j\equiv \vec x_{ij}$ downstream 
particles are identified by the condition $\vec n_i \cdot \vec x_{ij}>0$,
whereas particles in the upstream regime satisfy
$\vec n_i \cdot \vec x_{ij}<0$.  

Hydrodynamic variables in the two states are evaluated by summing the 
contribution of the neighboring particles  to the up- and downstream
 regimes. This is done by adopting a twofold weighting scheme.
The first $(F)$ weights the particle $j$ according to its projected distance 
from the up- and downstream positions:

   \begin{equation}
 U^F_{ij}= \zeta-|\vec n_i \cdot  \vec x_{ij}|/h_i~.
    \label{uf.eq}
   \end{equation}

The second $(S)$ weights the contribution of particle $j$ inversely
with its distance from the shock normal:

   \begin{equation}
 U^S_{ij}= \sqrt{ \vec x_{ij}^2-(\vec n_i \cdot \vec x_{ij})^2}/h_i~.
    \label{us.eq}
   \end{equation}

An estimator of the hydrodynamic variables $Z=\{\rho,~P,~c_s, ~\vec v\}$ 
is then given by

   \begin{equation}
 Z^s_i=   \frac{ \sum_{j \in s} w_{ij} Z_j}{ \sum_{j \in s}  w_{ij}}~,
   \label{zw.eq}
   \end{equation}

where $s=\{u,d\}$ is just a shorthand notation to indicate the two possible
states and

   \begin{equation}
w_{ij}=   m^2_j W(U_{ij}^F) W(U_{ij}^S)
   \label{wu.eq}
   \end{equation}

is the total weight.

In order to compute the shock Mach number we must now apply the flux 
conservation
laws between the up- and downstream hydrodynamic variables. 
However, unlike in Beck et al. (2016b), we allow here for the presence of bulk
 motions, such as those present in the cluster ICM. 

 We define the shock velocity in the lab frame to be 
$ \vec V^{lab}_{sh} = V_{sh} \vec n_{sh} $, and
in the shock rest frame   the shock mass flux conservation is 

   \begin{equation}
 \rho_u ( \vec V_u \cdot \vec n_{sh})  \equiv \rho_u V_u =
\rho_d ( \vec V_d \cdot \vec n_{sh})  \equiv \rho_d V_d~,
   \label{sha.eq}
   \end{equation}

where upstream and downstream velocities are measured in the shock rest frame.
This equation in the lab frame is \citep{sc20}

   \begin{equation}
 \rho_u ( \vec V^{lab}_u - V^{lab} _{sh}  \vec n_{sh})\cdot \vec n_{sh} = 
 \rho_d ( \vec V^{lab}_d - V^{lab} _{sh}  \vec n_{sh})\cdot \vec n_{sh}~,
   \label{shb.eq}
   \end{equation}

 and the shock  velocity in the lab frame is then the given by

   \begin{equation}
 V^{lab}_{sh}= \frac{ \vec n_{sh} \cdot \Delta ({\rho \vec V^{lab})}}
{ \Delta {\rho} } =\frac{(\rho_d V^{lab}_d-\rho_u V^{lab}_u)} {\rho_d-\rho_u}~.
   \label{shc.eq}
   \end{equation}

We  now define the particle Mach number $M_i$ as 

 \begin{equation}
M_i= \{ V^{lab}_{sh}/c^u_s  \}_i,
 \label{she.eq}
 \end{equation}
where we have used for particle $i$ the quantities previously derived 
from Equation (\ref{zw.eq}).

Finally, as in Beck et al. (2016b) , we apply a set of filterings criteria 
to avoid false detections and to reduce particle noise. 
We first require $P_i^d>(1+\epsilon) P_i^u$ and 
$\rho_i^d>(1+\epsilon) \rho_i^u$, where 
$\epsilon=0.05$. 
Moreover, the velocity divergence 

   \begin{equation}
 \Delta v_i= \vec n_{sh} \cdot ( \vec V_d- \vec V_u)=
  V_d- V_u~,
   \label{shd.eq}
   \end{equation}

must satisfy $\Delta v_i<0$ , since the shock compression ratio 
 $r=\rho_d/\rho_u$ is greater than unity and from Equation (\ref{sha.eq}) 
 one has $V_d< V_u$.
Additionally,  for the viscosity limiter  
  $f_i={|\vec \nabla \cdot \vec v|_i}/
  {(|\vec \nabla \cdot \vec v|_i+|\vec \nabla \times \vec v|_i)} $ 
 we set  a  threshold value for $f_i>0.6$  to avoid false
detections in shear flows.
 For all the gas particles  we initialize $M_i$ to zero, so that only those
particles which satisfy the filtering criteria have a non-zero Mach 
number.

%
%


\end{document}